\def\ref#1{\lbrack #1\rbrack}
\def\eck#1{\left\lbrack #1 \right\rbrack}
\def\eckk#1{\bigl[ #1 \bigr]}
\def\rund#1{\left( #1 \right)}
\def\abs#1{\left\vert #1 \right\vert}

\def\ave#1{\left\langle #1 \right\rangle}

\def\part#1#2{{\partial #1\over\partial #2}}

\def\Re{{\cal R}\hbox{e}}

\def\D{{\cal D}}

\def\d{{\rm d}}

\def\eps{{\epsilon}}

\def\arcminf {\hbox{$.\!\!^{\prime}$}}
\def\vp{\varphi}

\def\Real{{\rm I\mathchoice{\kern-0.70mm}{\kern-0.70mm}{\kern-0.65mm}%
  {\kern-0.50mm}R}}
\def\C{\rm C\kern-.42em\vrule width.03em height.58em depth-.02em
       \kern.4em}
\font \bolditalics = cmmib10
\def\bx#1{\leavevmode\thinspace\hbox{\vrule\vtop{\vbox{\hrule\kern1pt
        \hbox{\vphantom{\tt/}\thinspace{\bf#1}\thinspace}}
      \kern1pt\hrule}\vrule}\thinspace}
\def\Rm#1{{\rm #1}}
\def \vc #1{{\textfont1=\bolditalics \hbox{$\bf#1$}}}
{\catcode`\@=11
\gdef\SchlangeUnter#1#2{\lower2pt\vbox{\baselineskip 0pt \lineskip0pt
  \ialign{$\m@th#1\hfil##\hfil$\crcr#2\crcr\sim\crcr}}}
}
\def\gtrsim{\mathrel{\mathpalette\SchlangeUnter>}}
\def\lesssim{\mathrel{\mathpalette\SchlangeUnter<}}

\def\ueber#1#2{{\setbox0=\hbox{$#1$}%
  \setbox1=\hbox to\wd0{\hss$\scriptscriptstyle #2$\hss}%
  \offinterlineskip
  \vbox{\box1\kern0.4mm\box0}}{}}

\def\bx#1{\leavevmode\thinspace\hbox{\vrule\vtop{\vbox{\hrule\kern1pt
        \hbox{\vphantom{\tt/}\thinspace{\bf#1}\thinspace}}
      \kern1pt\hrule}\vrule}\thinspace}

\def\SFB{{This work was supported by the ``Sonderforschungsbereich
375-95 f\"ur
Astro--Teil\-chen\-phy\-sik" der Deutschen For\-schungs\-ge\-mein\-schaft.}}
\magnification=\magstep1
\input epsf
\voffset= 0.0 true cm
\vsize=19.8 cm     
\hsize=13.5 cm
\hfuzz=2pt
\tolerance=500
\abovedisplayskip=3 mm plus6pt minus 4pt
\belowdisplayskip=3 mm plus6pt minus 4pt
\abovedisplayshortskip=0mm plus6pt
\belowdisplayshortskip=2 mm plus4pt minus 4pt
\predisplaypenalty=0
\footline={\tenrm\ifodd\pageno\hfil\folio\else\folio\hfil\fi}

\def\la{\mathrel{\hbox{\rlap{\hbox{\lower4pt\hbox{$\sim$}}}\hbox{$<$}}}}
\def\ga{\mathrel{\hbox{\rlap{\hbox{\lower4pt\hbox{$\sim$}}}\hbox{$>$}}}}

\def\arcmin{\hbox{$^\prime$}}

\def\utw{\smash{\rlap{\lower5pt\hbox{$\sim$}}}}
\def\udtw{\smash{\rlap{\lower6pt\hbox{$\approx$}}}}

\def\getsto{\mathrel{\hbox{\rlap{$\gets$}\hbox{\raise2pt\hbox{$\to$}}}}}
\def\lid{\mathrel{\hbox{\rlap{\hbox{\lower4pt\hbox{$=$}}}\hbox{$<$}}}}
\def\gid{\mathrel{\hbox{\rlap{\hbox{\lower4pt\hbox{$=$}}}\hbox{$>$}}}}
\def\sol{\mathrel{\hbox{\rlap{\hbox{\raise4pt\hbox{$\sim$}}}\hbox{$<$}}}
}
\def\sog{\mathrel{\hbox{\rlap{\hbox{\raise4pt\hbox{$\sim$}}}\hbox{$>$}}}
}
\def\lse{\mathrel{\hbox{\rlap{\hbox{\raise4pt\hbox{$<$}}}\hbox{$\simeq$}
}}}
\def\gse{\mathrel{\hbox{\rlap{\hbox{\raise4pt\hbox{$>$}}}\hbox{$\simeq$}
}}}
\def\grole{\mathrel{\hbox{\lower2pt\hbox{$<$}}\kern-8pt
\hbox{\raise2pt\hbox{$>$}}}}
\def\leogr{\mathrel{\hbox{\lower2pt\hbox{$>$}}\kern-8pt
\hbox{\raise2pt\hbox{$<$}}}}
\def\loa{\mathrel{\hbox{\rlap{\hbox{\lower4pt\hbox{$\approx$}}}\hbox{$<$
}}}}
\def\goa{\mathrel{\hbox{\rlap{\hbox{\lower4pt\hbox{$\approx$}}}\hbox{$>$
}}}}

%
%

\font\kleinhalbcurs=cmmib10 scaled 833
\font\eightrm=cmr8
\font\sixrm=cmr6
\font\eighti=cmmi8
\font\sixi=cmmi6
\skewchar\eighti='177 \skewchar\sixi='177
\font\eightsy=cmsy8
\font\sixsy=cmsy6
\skewchar\eightsy='60 \skewchar\sixsy='60
\font\eightbf=cmbx8
\font\sixbf=cmbx6
\font\eighttt=cmtt8
\hyphenchar\eighttt=-1
\font\eightsl=cmsl8
\font\eightit=cmti8

\font\bxf=cmbx10
  \mathchardef\Gamma="0100
  \mathchardef\Delta="0101
  \mathchardef\Theta="0102
  \mathchardef\Lambda="0103
  \mathchardef\Xi="0104
  \mathchardef\Pi="0105
  \mathchardef\Sigma="0106
  \mathchardef\Upsilon="0107
  \mathchardef\Phi="0108
  \mathchardef\Psi="0109
  \mathchardef\Omega="010A
\def\rahmen#1{\vskip#1truecm}
\def\begfig#1cm#2\endfig{\par
\setbox1=\vbox{\rahmen{#1}#2}%
\dimen0=\ht1\advance\dimen0by\dp1\advance\dimen0by5\baselineskip
\advance\dimen0by0.4true cm
\ifdim\dimen0>\vsize\pageinsert\box1\vfill\endinsert
\else
\dimen0=\pagetotal\ifdim\dimen0<\pagegoal
\advance\dimen0by\ht1\advance\dimen0by\dp1\advance\dimen0by1.4true cm
\ifdim\dimen0>\vsize
\topinsert\box1\endinsert
\else\vskip1true cm\box1\vskip4true mm\fi
\else\vskip1true cm\box1\vskip4true mm\fi\fi}
\def\figure#1#2{\smallskip\setbox0=\vbox{\noindent\petit{\bf Fig.\ts#1.\
}\ignorespaces #2\smallskip
\count255=0\global\advance\count255by\prevgraf}%
\ifnum\count255>1\box0\else
\centerline{\petit{\bf Fig.\ts#1.\ }\ignorespaces#2}\smallskip\fi}

\def\xfigure#1#2#3#4{\midinsert\noindent
    $$\epsfxsize=#4truecm\epsffile{#3}$$
    \figure{#1}{#2}\endinsert}

\def\tabcap#1#2{\smallskip\vbox{\noindent\petit{\bf Table\ts#1\unskip.\
}\ignorespaces #2\smallskip}}
\def\begtab#1cm#2\endtab{\par
\ifvoid\topins\midinsert\vbox{#2\rahmen{#1}}\endinsert
\else\topinsert\vbox{#2\kern#1true cm}\endinsert\fi}
\def\rahmen#1{\vskip#1truecm}
\def\begpet{\vskip6pt\bgroup\petit}
\def\endpet{\vskip6pt\egroup}
\def\begref{\par\bgroup\petit
\let\it=\rm\let\bf=\rm\let\sl=\rm\let\INS=N}
\def\petit{\def\rm{\fam0\eightrm}%
\textfont0=\eightrm \scriptfont0=\sixrm \scriptscriptfont0=\fiverm
 \textfont1=\eighti \scriptfont1=\sixi \scriptscriptfont1=\fivei
 \textfont2=\eightsy \scriptfont2=\sixsy \scriptscriptfont2=\fivesy
 \def\it{\fam\itfam\eightit}%
 \textfont\itfam=\eightit
 \def\sl{\fam\slfam\eightsl}%
 \textfont\slfam=\eightsl
 \def\bf{\fam\bffam\eightbf}%
 \textfont\bffam=\eightbf \scriptfont\bffam=\sixbf
 \scriptscriptfont\bffam=\fivebf
 \def\tt{\fam\ttfam\eighttt}%
 \textfont\ttfam=\eighttt
 \normalbaselineskip=9pt
 \setbox\strutbox=\hbox{\vrule height7pt depth2pt width0pt}%
 \normalbaselines\rm
\def\vec##1{\setbox0=\hbox{$##1$}\hbox{\hbox
to0pt{\copy0\hss}\kern0.45pt\box0}}}%
\let\ts=\thinspace
%
\font \tafontt=     cmbx10 scaled\magstep2
\font \tafonts=     cmbx7  scaled\magstep2
\font \tafontss=     cmbx5  scaled\magstep2
\font \tamt= cmmib10 scaled\magstep2
\font \tams= cmmib10 scaled\magstep1
\font \tamss= cmmib10
\font \tast= cmsy10 scaled\magstep2
\font \tass= cmsy7  scaled\magstep2
\font \tasss= cmsy5  scaled\magstep2
\font \tasyt= cmex10 scaled\magstep2
\font \tasys= cmex10 scaled\magstep1
\font \tbfontt=     cmbx10 scaled\magstep1
\font \tbfonts=     cmbx7  scaled\magstep1
\font \tbfontss=     cmbx5  scaled\magstep1
\font \tbst= cmsy10 scaled\magstep1
\font \tbss= cmsy7  scaled\magstep1
\font \tbsss= cmsy5  scaled\magstep1

\newbox\chsta\newbox\chstb\newbox\chstc
\def\centerpar#1{{\advance\hsize by-2\parindent
\rightskip=0pt plus 4em
\leftskip=0pt plus 4em
\parindent=0pt\setbox\chsta=\vbox{#1}%
\global\setbox\chstb=\vbox{\unvbox\chsta
\setbox\chstc=\lastbox
\line{\hfill\unhbox\chstc\unskip\unskip\unpenalty\hfill}}}%
\leftline{\kern\parindent\box\chstb}}
 \def \chap#1{
    \vskip24pt plus 6pt minus 4pt
    \bgroup
 \textfont0=\tafontt \scriptfont0=\tafonts \scriptscriptfont0=\tafontss
 \textfont1=\tamt \scriptfont1=\tams \scriptscriptfont1=\tamss
 \textfont2=\tast \scriptfont2=\tass \scriptscriptfont2=\tasss
 \textfont3=\tasyt \scriptfont3=\tasys \scriptscriptfont3=\tenex
     \baselineskip=18pt
     \lineskip=18pt
     \raggedright
     \pretolerance=10000
     \noindent
     \tafontt
     \ignorespaces#1\vskip7true mm plus6pt minus 4pt
     \egroup\noindent\ignorespaces}%
 \def \sec#1{
     \vskip25true pt plus4pt minus4pt
     \bgroup
 \textfont0=\tbfontt \scriptfont0=\tbfonts \scriptscriptfont0=\tbfontss
 \textfont1=\tams \scriptfont1=\tamss \scriptscriptfont1=\kleinhalbcurs
 \textfont2=\tbst \scriptfont2=\tbss \scriptscriptfont2=\tbsss
 \textfont3=\tasys \scriptfont3=\tenex \scriptscriptfont3=\tenex
     \baselineskip=16pt
     \lineskip=16pt
     \raggedright
     \pretolerance=10000
     \noindent
     \tbfontt
     \ignorespaces #1
     \vskip12true pt plus4pt minus4pt\egroup\noindent\ignorespaces}%
 \def \subs#1{
     \vskip15true pt plus 4pt minus4pt
     \bgroup
     \bxf
     \noindent
     \raggedright
     \pretolerance=10000
     \ignorespaces #1
     \vskip6true pt plus4pt minus4pt\egroup
     \noindent\ignorespaces}%
 \def \subsubs#1{
     \vskip15true pt plus 4pt minus 4pt
     \bgroup
     \bf
     \noindent
     \ignorespaces #1\unskip.\ \egroup
     \ignorespaces}
\def\footnoterule{\kern-3pt\hrule width 2true cm\kern2.6pt}
\newcount\footcount \footcount=0
\def\advftncnt{\advance\footcount by1\global\footcount=\footcount}
\def\fonote#1{\advftncnt$^{\the\footcount}$\begingroup\petit
       \def\textindent##1{\hang\noindent\hbox
       to\parindent{##1\hss}\ignorespaces}%
\vfootnote{$^{\the\footcount}$}{#1}\endgroup}

\newcount\sterne
\outer\def\byebye{\bigskip\typeset
\sterne=1\ifx\speciali\undefined\else
\bigskip Special caracters created by the author
\loop\smallskip\noindent special character No\number\sterne:
\csname special\romannumeral\sterne\endcsname
\advance\sterne by 1\global\sterne=\sterne
\ifnum\sterne<11\repeat\fi
\vfill\supereject\end}
\def\typeset{\centerline{\petit This article was processed by the author
using the \TeX\ Macropackage from Springer-Verlag.}}
\def\s{{\rm(s)}}
\chap{Detection of (dark) matter concentrations via weak gravitational
lensing} 
\medskip
\centerline{\bf Peter Schneider}
\centerline{\bf Max-Planck-Institut f\"ur Astrophysik}
\centerline{\bf Postfach 1523}
\centerline{\bf D-85740 Garching, Germany}
\bigskip
\sec{Abstract}
The distortion of images of faint background galaxies by (weak)
gravitational lensing can be used to measure the mass distribution of
the deflector. Reconstruction methods for the mass profile of lensing
clusters have been developed and successfully tested. Alternatively,
the image distortions can be used to define a weighted mean of the
mass inside a circular aperture, as was first suggested by
Kaiser. This ``aperture mass'' has the advantage that strict error bars
can be obtained from the data itself, and that a strict lower limit of
the lens mass inside a circle can be obtained. 

The aperture mass can thus be used to {\it detect} dark matter
concentrations. Keeping in mind that wide-field cameras will become
increasingly available, this method can be used to search for mass
concentrations on wide-field images. To do this,
the aperture mass measure is generalized to account for
different weighting functions. For each such weighting function, a
signal-to-noise ratio can be calculated. For an assumed mass profile
of the density concentrations, 
the weighting function can be chosen such as to
maximize the resulting signal-to-noise ratio. Assuming that dark
matter halos can be approximated by an isothermal profile over a large
range of radius, a weighting function is constructed which is adapted
to this density profile and which yields a smooth signal-to-noise
map. Numerical simulations which adopt parameters characteristic of
4-m class telescopes are then used to show that dark halos with
a velocity dispersion in excess of $\sim 600$\ts km/s can be reliably
detected as significant peaks in the signal-to-noise map. The effects
of seeing and an anisotropic PSF are then investigated and shown to be
less important than might be feared. It is thus suggested that the
method of aperture mass measures developed here can be used to obtain a
mass-selected sample of dark halos, in contrast to flux-selected
samples. Shear fields around high-redshift bright QSOs as detected by
Fort et al. provide a first successful application of this
strategy. The simplicity of the method allows its routine application
to wide-field images of sufficient depth and image quality.

\vfill\eject

\sec{1 Introduction}
Objects in the universe are usually discovered by means of their
emitted radiation -- the brighter the objects are, the easier they are
detected. Therefore, all samples of objects are flux limited. 
The inclusion of objects thus depends, among other things, on the
wavelength of observation. As an example, clusters of galaxies can be
selected either by X-ray surveys or from optical images. There is no
reason to expect that the clusters selected by these two methods will
share most of their properties; in the former case, clusters with a
dense intracluster medium will be preferentially included in the
samples, whereas optical survey are sensitive to clusters which
contain many bright galaxies. Given that the intracluster gas and the
galaxies constitute only a small fraction of the cluster mass, which
is the quantity cosmologists are most interested in, it is by no means
clear that flux-limited samples reveal a fair representation of (dark)
matter concentrations in the universe.

Gravitational lensing offers the possibility to detect `unseen' matter
concentrations just from their gravitational effect on light rays. For
example, the quantitative analysis of the currently existing surveys for
gravitationally-lensed QSOs can rule out a significant density of
compact lenses in the universe with mass $\gtrsim 10^{11} M_\odot$
(e.g., Kochanek 1993, 1995; Maoz \& Rix 1993). Constraints from QSO
variability and spectral characteristics have yielded strict upper
limits on the cosmological density of compact objects with
$10^{-3}M_\odot \lesssim M\lesssim 10^2M_\odot$ (e.g., Schneider 1993;
Dalcanton et al.\ 1994). Since most groups and clusters of galaxies
are expected not to be sufficiently compact to yield multiply-imaged
QSO images, these surveys do not yield significant constraints on
their cosmic number density.

On the other hand, the mass distribution in clusters can be
investigated with gravitational lensing, either from studying arcs and
arclets (see the recent review by Fort \& Mellier 1994), or from the
`weak' distortion of numerous faint background galaxies by the tidal
gravitational field of the cluster (Tyson, Valdes \& Wenk 1990; Kaiser
\& Squires 1993). These techniques have already yielded most
interesting results; e.g., the core radius of the dark matter in
clusters was found to be considerably smaller than that normally 
concluded from
X-ray studies (e.g., Miralda-Escud\'e \& Babul 1995 
and references therein), and the
mass-to-light ratio in the cluster MS1224 was found to be considerably
larger than `typical' estimates in other clusters (Fahlman et al.\ts
1994). This object 
supports the possibility that clusters of rather different
mass-to-light ratios exist, and that optical survey are (strongly)
biased towards the detection of only the brightest subset of these
clusters. 

These lensing studies of clusters have been performed exclusively on
clusters which were known before from optical or X-ray
surveys. However, Fort et al.\ts (1996) have recently started to
develop a strategy for the detection of mass concentrations which are
not included in existing samples. Supported by the findings that there
is a statistically significant overdensity of (foreground) galaxies
and clusters around high-redshift bright QSOs (Fugmann 1990;
Bartelmann \& Schneider 1993, 1994; Rodrigues-Williams \& Hogan 1994;
Benitez \& Martinez-Gonzalez 1995; Hutchings 1995; Seitz \& Schneider
1995b), on scales in excess of several arcminutes, they assumed that
these apparently most luminous QSOs are affected by the magnification
bias. This magnification cannot be due to single galaxies, as the
relevant scale in this case would be of order arcseconds, but must
come from more extended mass concentrations. Fort et al.\ts (1996)
have tried to discover the presence of (foreground) mass
concentrations, i.e., by mapping the (weak) shear in the field of the
QSO, by investigating the systematic ellipticity component of
background galaxies. In several of the fields which they imaged, a
significant shear field was detected, which provides a direct
verification of the overdensity of material near the line-of-sight to
these QSOs.  One might suspect that in this way, one could obtain a
`mass-selected' sample of groups and/or clusters of galaxies.

The main drawback of this method is that high-redshift, bright QSOs
are rare, and these mass concentrations are selected only along these
rare lines-of-sight. On the other hand, several wide-field imaging
cameras are either installed or planned, and the resulting images can
be used to systematically search for dark matter concentrations.
In this paper, a statistical method for the detection of (dark) matter
concentrations is described which is also based on weak lensing of
background galaxies. From the image ellipticities of background
galaxies, one can derive a shear map, and from that, the surface mass
density of the lens can be reconstructed (Kaiser \& Squires 1993;
Schneider \& Seitz 1995; Seitz \& Schneider 1995a, 1996b; Bartelmann et
al.\ts 1996). One could then analyze the resulting mass maps to search
for matter concentrations. However, the relation between galaxy
ellipticities and the resulting mass map is quite complicated, and no
simple test for a statistical significance of a concentration has been
derived yet (although numerical simulations can in principle derive
such significance levels). Alternatively, Kaiser (1995) pointed out
that the mean surface mass density in a circular aperture can be
directly obtained from the shear,
$$
\zeta(x_1,x_2):=\bar\kappa(x_1)-\bar\kappa(x_1,x_2)
={2 x_2^2\over x_2^2-x_1^2}\int_{x_1}^{x_2}{\d x\over x}
\ave{\gamma_{\rm t}}(x)\quad ,
\eqno (1)
$$
where $\bar\kappa(x_1)$ is the mean dimensionless surface mass density
within a circle of radius $x_1$ around a point, $\bar\kappa(x_1,x_2)$
is the mean dimensionless surface mass density in an annulus of radii
$x_1$ and $x_2>x_1$ around the same point, and $\ave{\gamma_{\rm t}}(x)$
is the mean tangential component of the shear at distance $x$ from
that reference point. Given that $\kappa$ is non-negative, the above
relation can be used to put a rigorous lower bound on the mass
contained inside the circle of radius $x_1$ around the reference
point.

In this paper, the relation (1) is first generalized to allow for
arbitrary weight functions. If one assumes that the mass
concentrations one expects to detect have an approximately isothermal
profile over a significant fraction of their extent, the weight
function can be optimized for the detection of such objects. The
important point to note then is that these aperture measures have 
well-defined and simple statistical properties. This allows to
generate a signal-to-noise map from the ellipticities of the faint
galaxies, in which sufficiently strong lenses will appear as
significant (say, $\gtrsim 4$-$\sigma$) peaks. This method is then
tested on synthetically generated data sets, and it is shown that it
lives up to the expectation. In particular, the signal-to-noise map is
not corrupted by small-scale deflectors (such as individual galaxies),
nor by larger-scale deflectors. Perhaps most importantly, the
signal-to-noise maps are much more weakly affected by seeing and an
anisotropic point-spread function than one might fear, at least as
long as the PSF is fairly stable over the `filter scale'. As will be
shown in Sect.\ts 4, `isothermal' mass concentrations with a velocity
dispersion in excess of $\sim 600$\ts km/s can be reliably detected by
our new method, where this lower limit depends on the seeing and the
depth of the observations. We therefore conclude that this detection
method, which is very easy to implement numerically and which is fast,
can be used routinely on wide-field images to search for dark mass
concentrations. 

\sec{2 Generalized aperture densitometry}
\subs{2.1 Notation}
The dimensionless surface mass density $\kappa(\vc x)$ as a
function of angular position $\vc\theta$ is related to the physical
surface mass density $\Sigma(\vc\theta)$ by
$$
\kappa(\vc x)={\Sigma(\vc x)\over\Sigma_{\rm cr}}\quad,
\eqno (2a)
$$
with the critical surface mass density 
$$
\Sigma_{\rm cr}={c^2 D_{\rm s}\over 4 \pi G D_{\rm d} D_{\rm
ds}}\quad,
\eqno (2b)
$$
where the $D$'s denote the distances to the lens and source, and from
the lens to the source (we use the same notation as in Schneider,
Ehlers \& Falco 1992). $\kappa$ is related to the
deflection potential $\psi(\vc x)$ by a Poisson-like equation,
$$
\nabla^2 \psi=2\kappa\quad ,
\eqno (3a)
$$
or
$$
\psi(\vc x)={1\over \pi}\int_{\Real^2}\d^2 x'\;
\kappa(\vc x')\,\ln\rund{\abs{\vc x-\vc x'}}\quad .
\eqno (3b)
$$
The shear $\gamma$ is a two-component quantity which describes the
traceless part of the Hessian of $\psi$; in particular, we write the
shear as a complex number $\gamma=\gamma_1+\Rm i \gamma_2$, with
$$
\gamma_1={1\over 2}\rund{\psi_{11}-\psi_{22}}\quad,\quad
\gamma_2=\psi_{12}\quad,
\eqno (4)
$$
where the indices on $\psi$ denote partial derivatives with respect to
$x_i$. Combining (3) and (4), one finds that
$$
\gamma(\vc x)={1\over \pi}\int_{\Real^2}\d^2 x'\;\D(\vc x-\vc x')
\kappa(\vc x')\quad ,
\eqno (5)
$$
where the complex kernel $\D$ is given by
$$
\D(\vc x)=- {x_1^2-x_2^2+2\Rm i x_1 x_2 \over \abs{\vc x}^4}\quad.
\eqno(6)
$$
If we write the vector $\vc x$ as a complex number $X:=x_1+\Rm i x_2$,
then 
$$
\D(\vc x)={-1\over (X^*)^2}\quad ,
\eqno (7)
$$
where the asterisk denotes complex conjugation. As was shown by Kaiser
\& Squires (1993), the relation (5) can be inverted to yield
$$
\kappa(\vc x)={1\over \pi} \Re\eck{\int_{\Real^2} \d^2 x'\;
\D^*(\vc x-\vc x')\,\gamma(\vc x')}+\kappa_0\quad ,
\eqno (8)
$$
where $\kappa_0$ is an arbitrary constant, and $\Re(X)$ denotes the
real part of the complex number $X$. Hence,  the
surface mass density can be determined from the shear field, up to an
overall additive constant.

\subs{2.2 Aperture measures}
Consider a point $\vc x_0$ and a weight function $w(x)$. We then
define the quantity
$$
m(\vc x_0):=\int\d^2 x\;\kappa(\vc x)\,w\rund{\abs{\vc x-\vc x_0}}
=\int\d^2 x\;\kappa(\vc x+\vc x_0)\,w\rund{\abs{\vc x}}\quad ,
\eqno (9)
$$
which is the integral of $\kappa$ in a circular aperture around $\vc
x_0$, weighted by the function $w$. Inserting (8), one finds
$$
m(\vc x_0)={1\over \pi}\Re\eck{\int_{\Real^2}\d^2 x'\;\gamma(\vc x')
\int\d^2 x\;\D^*\rund{\vc x+\vc x_0-\vc x'}\,w\rund{\abs{\vc x}}}
+\int\d^2 x\;\kappa_0\,w\rund{\abs{\vc x}}\quad .
\eqno (10)
$$
The quantity $m$ does not depend on the undetermined constant
$\kappa_0$ if one requires that
$$
\int \d x\;x\,w(x)=0\quad .
\eqno (11)
$$
Then, by changing variables in (10) to $\vc y=\vc x'-\vc x_0$, and
inserting the definition (7) of $\D$, one obtains
$$
m(\vc x_0)={-1\over \pi}\Re\eck{\int_{\Real^2}\d^2 y\;\gamma(\vc y+\vc x_0)
\int_0^\infty\d x\;x\,w(x)\int_0^{2\pi}{\d\vp\over\rund{X-Y}^2}}\quad,
\eqno (12)
$$
where we have again used capital letters to denote complex number,
i.e., $Y=y_1+\Rm i y_2$. The final
integral in (12) can be evaluated by substituting $X=x\Rm e^{\Rm
i\vp}$, so that $\d\vp=-\Rm i\, \d X/X$, and the integral is transformed
into a loop integral. The resulting integrand then has two poles, one
at $X=0$, and one at $X=Y$. Depending on whether $x<\abs{Y}$ or not,
the second pole lies inside or outside the loop. Application of the
residue theorem then yields
$$
\int_0^{2\pi}{\d\vp\over\rund{X-Y}^2}
={\pi\over Y^2}\eck{2 \Rm H\rund{\abs{X}-\abs{Y}}
-\abs{Y}\delta\rund{\abs{X}-\abs{Y}}}\quad ,
\eqno (13)
$$
where $\Rm H(x)$ is the Heaviside step function, and $\delta(x)$ is
Dirac's delta `function'. Defining the {\it tangential shear
$\gamma_{\rm t}(\vc y;\vc x_0)$ at
position $\vc y$ relative to the point $\vc x_0$} by
$$
{\gamma_{\rm t}(\vc y;\vc x_0)\over \abs{Y}^2}
=-\Re\rund{\gamma(\vc y+\vc x_0)\over Y^2}\quad,
\eqno (14)
$$
one obtains from combining (12)--(14)
$$
m(\vc x_0)=\int\d^2 y\;{\gamma_{\rm t}(\vc y;\vc x_0)
\over \abs{\vc y}^2} \,Q(\abs{\vc y})\quad ,
\eqno (15)
$$
where we have defined
$$
Q(x):=2\int_0^x\d x\; x\,w(x) -x^2\,w(x)\quad .
\eqno (16)
$$
Equation (15) states that the
$w$-weighted integral of the surface mass density can be expressed as
an integral over the tangential shear, weighted by the function
$Q$. This result was first derived by Kaiser et al.\ts (1994; see also
the recent preprint by Squires and Kaiser 1996 for a generalization to
non-circular apertures), using Gauss' law.
To see the usefulness of this relation, consider the case that
$w(x)$ is chosen such that 
$$
w(x)=0\quad{\rm for}\quad  x>R\quad ; 
\eqno (17)
$$
then, owing to the condition (11), $Q(x)=0$ for $x>R$. Thus, in this
case, the quantity $m(\vc x_0)$ can be expressed as an integral of the
tangential shear {\it over a finite circle} around $\vc
x_0$. Furthermore, if $w(x)$ is chosen such that $w(x)=$\ts const. for
$x\le x_1<R$, then $Q(x)=0$ for $x\le x_1$. Therefore, $m(\vc x_0)$
can then be expressed as an integral of the tangential shear over the
annulus $x_1\le\abs{\vc x-\vc x_0}\le R$. The $\zeta$-statistics (1)
derived by Kaiser (1995) is a special case (15), which is obtained by
setting $w=\hat w_{\rm K}$, with (setting $x_2=R$)
$$\eqalign{
\hat w_{\rm K}(x)&={1\over \pi \,x_1^2}
\quad{\rm for}\quad 0\le x < x_1\;,\cr
\hat w_{\rm K}(x)&={-1\over \pi (R^2-x_1^2)}
\quad{\rm for}\quad x_1\le x < R\;,\cr
\hat w_{\rm K}(x)&=0\quad {\rm for}\quad x\ge R\quad .\cr}
\eqno (18)
$$
The obvious advantage of the $\zeta$-statistics (1) is that it can be
used to directly infer a lower bound on the mass inside a circle of
radius $x_1$ around a point $\vc x_0$, due to the non-negativity of
the surface mass density, i.e., $\bar\kappa(x_1,x_2)\ge 0$. The reason
why we are interested in the generalized 
aperture mass measure $m$ is that the
weight function $w$ can be adjusted such that the $m$-statistics is
`optimized' for the detection of mass overdensities, as will be shown
in the next section.

\sec{3 Signal-to-noise statistics}
The observational estimate of the shear field $\gamma$ proceeds by
investigating the distortion of the images of faint background
galaxies. Using the same notation as in Schneider \& Seitz (1995) and
Schneider (1995), except that here we have defined the sign of
$\gamma$ differently as in the quoted papers, we denote by $\eps^\s$
the complex ellipticity of a source, such that for an elliptical
source with axis ratio $r\le 1$, $\abs{\eps^\s}=(1-r)/(1+r)$, and the
phase of $\eps^\s$ is twice the angle between the major axis and the
positive $x_1$-direction (see also Sect.\ts 4.4 below). The (complex)
image ellipticity $\eps$ is 
obtained from $\eps^\s$ and the local lens parameters as
$$
\eps={\eps^\s + g\over 1+g^* \eps^\s}\quad,
\eqno (19)
$$
where $g=\gamma/(1-\kappa)$ is the (complex) {\it reduced
shear}.\fonote{The transformation (19) is valid only if $\abs{g}<1$;
if $\abs{g}\ge 1$, a different transformation applies (see Seitz \&
Schneider 1996a). However, in this paper we shall concentrate mainly
on non-critical lenses, for which (19) is valid.} In
the case of weak lensing, which should mainly be considered here, $g\approx
\gamma$. Assuming that the intrinsic orientation of the sources is
random, then the expectation value of the average ellipticity of a
local sample of galaxy images becomes (see Schramm \& Kayser 1995;
Seitz \& Schneider 1996a)
$$
\ave{\eps}_\eps=g\approx \gamma\quad .
\eqno (20)
$$
Thus, the ellipticity of a galaxy image is an unbiased estimate of the
local shear in the case of weak lensing.

Let $n$ be the number density of galaxy images, and let $\vc x_i$ and
$\eps_i$ denote the position vectors and ellipticities of the
galaxies. Defining in analogy to (14) the {\it tangential ellipticity
$\eps_{{\rm t}i}(\vc x_0)$ of galaxy image $i$ relative to the
point $\vc x_0$} by
$$
\eps_{{\rm t}i}(\vc x_0)=-\Re\rund{\eps_i (X_i-X_0)^*\over 
(X_i-X_0)}\quad ,
\eqno (21)
$$
where again complex number representation of the vectors $\vc x_i$ and
$\vc x_0$ was used on the right-hand-side, we can write down a
discrete version of (15) as
$$
m(\vc x_0)={1\over n}\sum_i {\eps_{{\rm t}i}(\vc x_0)\over
\abs{\vc x_i-\vc x_0}^2}\,Q\rund{\abs{\vc x_i-\vc x_0}}\quad ,
\eqno (22)
$$
where the sum extends over all galaxy images within the aperture,
i.e., within the region where $Q\ne 0$. 

In the case of no lensing, the expectation value of $m$ vanishes,
$\ave{m}_\eps =0$, because the expectation value of the tangential
ellipticity is zero. The dispersion $\sigma_{m\rm d}$ of $m$ can be
calculated by squaring (22) and taking the expectation value, which
leads to 
$$
\sigma_{m\rm d}^2(\vc x_0)={\sigma_\eps^2 \over 2 n^2}
\sum_i {Q^2\rund{\abs{\vc x_i-\vc x_0}}\over \abs{\vc x_i-\vc x_0}^4}\quad,
\eqno (23)
$$
where the fact was used that the expectation value of the product of
tangential ellipticities of different images vanishes,
$\ave{\eps_{{\rm t}i} \eps_{{\rm t}j}}=0$ for $i\ne j$, since the
orientation of galaxy images are assumed to be mutually unrelated; 
$\sigma_\eps$ is the dispersion of the intrinsic ellipticity
distribution, $\ave{\abs{\eps}^2}=\ave{\abs{\eps^s}^2}=\sigma_\eps^2$,
and the factor $1/2$ is due to the fact that the dispersion in one
component of the ellipticity is $1/\sqrt{2}$ times the dispersion of
the complex ellipticity, $\ave{\eps_{{\rm t}i} \eps_{{\rm
t}i}}=\sigma_\eps^2/2$. Thus, we can define {\it the signal-to-noise
ratio} $S$ at position $\vc x_0$ as
$$
S(\vc x_0)={\sqrt{2}\over \sigma_\eps}
{\sum_i {\eps_{{\rm t}i}(\vc x_0)\over
\abs{\vc x_i-\vc x_0}^2}\,Q\rund{\abs{\vc x_i-\vc x_0}} \over
\sqrt{ \sum_i {Q^2\rund{\abs{\vc x_i-\vc x_0}}\over 
\abs{\vc x_i-\vc x_0}^4}}} \quad.
\eqno (24)
$$
Note that $S$ is independent of the normalization of $Q$, and thus the
normalization of $w$.

Up to now the weight function $w(x)$, and thus $Q(x)$, has been
undetermined. We shall now derive the shape of $w(x)$ which maximizes
the signal-to-noise ratio for a given mass profile. In order to
simplify the notation, we set $\vc x_0=\vc 0$ in the following
consideration. The expectation value of $m$ is
$$
\ave{m}_{\rm d}={1\over n}\sum_i {\gamma_{\rm t}(\vc x_i)\over
\abs{\vc x_i}^2}\,Q(\abs{\vc x_i})\quad .
\eqno (25)
$$
Assuming that $w(x)=0$ for $x>R$, so that $Q(x)=0$ for $x>R$ (we shall
call $R$ the `aperture size', or the `filter scale'), we can
perform an ensemble average by averaging (25) over the probability
distribution for the galaxy positions,
$$
\ave{m}_{\rm c}=\eck{\prod_{k=1}^N \int{\d^2 x_k\over \pi
R^2}}\ave{m}_{\rm d}
=2\pi \int_0^R \d x\;x\,\ave{\gamma_{\rm t}}(x) {Q(x)\over x^2} ,\quad
\eqno (26)
$$
where $N$ is the expected number of galaxies within $R$, $N=n\pi R^2$,
and $\ave{\gamma_{\rm t}}(x)$ is the mean tangential shear on a circle of
radius $x$ around the point $\vc x_0=\vc 0$. The indices `c' and `d'
stand for continuous and discrete.
Accordingly, we can perform a
similar ensemble average of the dispersion (23), to obtain
$$
\sigma_{m\rm c}^2=\eck{\prod_{k=1}^N \int { \d^2 x_k\over \pi R^2}}
\sigma_{m\rm d}^2
={\pi \sigma_\eps^2\over n}\int_0^R\d x\;{Q^2(x)\over x^3}\quad .
\eqno (27)
$$
From (26) and (27) one can define the ensemble-averaged
signal-to-noise ratio $S_{\rm c}$ as
$$
S_{\rm c}={2\sqrt{\pi n}\over \sigma_\eps}
{\int_0^R \d x\; \ave{\gamma_{\rm t}}(x) Q(x)/x \over
\sqrt{\int_0^R \d x\; Q^2(x)/x^3}}\quad .
\eqno (28)
$$
According to the Cauchy--Schwarz inequality, 
$$
\rund{\int\d x\;f(x)\,g(x)}^2\le \rund{\int\d x\;f^2(x)}
\rund{\int\d x\;g^2(x)} \quad,
$$
with equality if and only if $f/g=$\ts const., one sees by identifying
$f=Q/x^{3/2}$ and $f g=\ave{\gamma_{\rm t}} Q/x$ that $S_{\rm c}^2$ is
maximized if
$$
Q(x)\propto x^2 \ave{\gamma_{\rm t}}(x)\quad .
\eqno (29)
$$
The corresponding weight function $w(x)$ can be obtained by
differentiation of (16), which yields a first-order ordinary
differential equation for $w$; the integration constant is fixed by
the requirement (11). The result is
$$
w(x)=2\int_x^R \d y\;{\ave{\gamma_{\rm t}}(y)\over y}
-\ave{\gamma_{\rm t}}(x)\quad .
\eqno (30)
$$
The result (30) for the weight function $w$ which maximizes the
signal-to-noise ratio for a given functional form of $\ave{\gamma_{\rm
t}}(y)$ can be written is a more suggestive form. For that, we use the
fact that 
$$
\bar \kappa(y)-\ave{\kappa}(y)=\ave{\gamma_{\rm t}}(y)\quad
\eqno (31)
$$
(see Bartelmann 1995). Here, $\bar\kappa(y)$ is the mean mass density
within a circle of radius $y$, and $\ave{\kappa}(y)$ is the mean mass
density on the circle with radius $y$. We can actually obtain this
result, which is trivially valid for axially-symmetric mass
distributions, by setting $w(x)=\Rm H(y-x)/(\pi y^2)-\delta(x-y)/(2\pi
y)$ in (9). Then, $m$ becomes the left-hand-side of Eq.\ts (31),
whereas by inserting the corresponding function $Q$ into (15), the
right-hand-side of (31) is obtained. Using that
$$
\bar\kappa(x)={2\over x^2}\int_0^x\d x'\;x' \ave{\kappa}(x')\quad ,
$$
one finds that
$$
\int_y^R\d x\;{\ave{\gamma_{\rm t}}(x)\over x}=
\int_y^R\d x\;{2\over x^3}\int_0^x \d z\; z\,\ave{\kappa}(z)
-\int_y^R \d x\;{\ave{\kappa}(x)\over x}\quad .
$$
After interchanging the order of integration, one finds from (30) 
after a short calculation that
$$
w(x)=\ave{\kappa}(x)-\bar\kappa(R)\quad .
\eqno (32)
$$
Correspondingly, one obtains, either from (32) and (16), or from (29)
and (31), that
$$
Q(x)=x^2\eckk{\bar\kappa(x)-\ave{\kappa}(y)}\quad .
\eqno (33)
$$
Whereas this `optimal' weight function maximizes the signal-to-noise
ratio, and should therefore be best suited for the purpose of this
paper, it is not practical to use it, for two reasons: first, the mass
profile of density inhomogeneities is not known a priori, and thus the
optimal weight function cannot be determined. Secondly, the weight
function is neither smooth nor continuous at $x=R$, and therefore the
$S$ statistics will not be a continuous function of position $\vc
x_0$. Since we want to search for significant peaks in the
$S$-function to identify mass inhomogeneities, this non-continuous
weight function will cause patterns in this
map which are difficult to interprete. We shall come back to this
point.
\medskip
%
\tabcap{1}{For the weight function $w$ as given in (34), the parameter
$\alpha$, $b$, and $c$ are given for several combinations of $\nu_1$
and $\nu_2$. In addition, $100 G$ is given, where $G=R^{-2}\int_0^R\d
y\,y^{-3}\,Q^2(y)$}
$$
\vbox{
\halign{\strut\offinterlineskip
        \tabskip=15pt
        \vrule#
        &$#$
        &\vrule#
        &$#$&\vrule#
        &$#$&\vrule#
        &\hfill$#$\hfill&\vrule#
        &#&\vrule#
        &#&\tabskip=0pt\vrule#\cr
\noalign{\hrule}
&\hfill \nu_1 \hfill&&\hfill\nu_2\hfill &&\hfill\alpha\hfill  &&b(\nu_2-1)^3 
&&\hfill c\hfill  &&\hfill 100 G\hfill&\cr
\noalign{\hrule}
&0.05  &&0.80  &&0.7133  &&0.1511  &&0.1239  &&0.7574&\cr
&0.05  &&0.85  &&0.7818  &&0.1450  &&0.1203  &&0.7667&\cr
&0.05  &&0.90  &&0.8528  &&0.1398  &&0.1170  &&0.7758&\cr
&0.05  &&0.95  &&0.9256  &&0.1351  &&0.1138  &&0.7845&\cr
&0.10  &&0.80  &&0.7148  &&0.3596  &&0.2555  &&2.4449&\cr
&0.10  &&0.85  &&0.7825  &&0.3435  &&0.2483  &&2.4844&\cr
&0.10  &&0.90  &&0.8531  &&0.3298  &&0.2415  &&2.5223&\cr
&0.10  &&0.95  &&0.9257  &&0.3180  &&0.2351  &&2.5588&\cr
&0.15  &&0.80  &&0.7171  &&0.6251  &&0.3844  &&4.8258&\cr
&0.15  &&0.85  &&0.7837  &&0.5941  &&0.3740  &&4.9213&\cr
&0.15  &&0.90  &&0.8535  &&0.5682  &&0.3642  &&5.0125&\cr
&0.15  &&0.95  &&0.9258  &&0.5461  &&0.3549  &&5.0997&\cr
&0.20  &&0.80  &&0.7201  &&0.9544  &&0.5050  &&7.8277&\cr
&0.20  &&0.85  &&0.7851  &&0.9022  &&0.4922  &&8.0142&\cr
&0.20  &&0.90  &&0.8541  &&0.8593  &&0.4800  &&8.1912&\cr
&0.20  &&0.95  &&0.9259  &&0.8232  &&0.4684  &&8.3591&\cr
\noalign{\hrule}
}}
$$

Nevertheless, the optimal weight function should provide a guide for
constructing weight functions which give rise to a smooth $S$-map. As
for the  weight function (18) used by Kaiser (1995), we want to have that
$Q(x)=0$ for $x\le x_1=:\nu_1 R$ -- for example, if the mass
overdensity is associated with bright galaxies, there will be no faint
galaxies visible in the very inner part. This implies that $w(x)=$\ts
const. for $x\le \nu_1 R$. Second, since one might expect that many of
the mass inhomogeneities will have approximately an isothermal
profile, corresponding to $\ave{\kappa}(x)\propto 1/x$, the weight
function should behave like $1/x-{\rm const.}$ over a large interval.
Third, in order to obtain a smooth function $S(\vc x_0)$, $w(x)$
should go to zero at $x=R$. The following form for $w(x)$ is
suggested:
$$\eqalign{
w(x)&=1\quad {\rm for}\quad x\in[0,\nu_1 R]\;,\cr
w(x)&={1\over 1-c}\rund{ {\nu_1 R\over \sqrt{ (x-\nu_1 R)^2+(\nu_1
R)^2}}-c} \quad {\rm for}\quad x\in [\nu_1 R,\nu_2 R]\;,\cr
w(x)&= {b\over R^3}(R-x)^2(x-\alpha R)
\quad{\rm for}\quad x\in[\nu_2 R,R]\;,\cr}
\eqno (34)
$$
where the constants $\alpha$, $b$ and $c$ are determined from
requireing continuity and smoothness (i.e., continuous first
derivative) at $x=\nu_2 R$, and by the normalization constraint
(11). Whereas these constants can be calculated analytically, the
resulting formulae are not very enlightening. For several combinations
of $\nu_1$ and $\nu_2$, they are tabulated in Table 1. The function
$w(x)$ is plotted for four different combinations of $\nu_1$
and $\nu_2$ in Fig.\ts 1.
\xfigure{1}{The weight function $w(x)$ as defined in (34) as a
function of $x/R$, for $(\nu_1,\nu_2)=(0.1,0.8)$ (solid curve), 
(0.1,0.9) (dotted curve), (0.2,0.8) (short-dashed curve) and (0.2,0.9)
(long-dashed curve)}{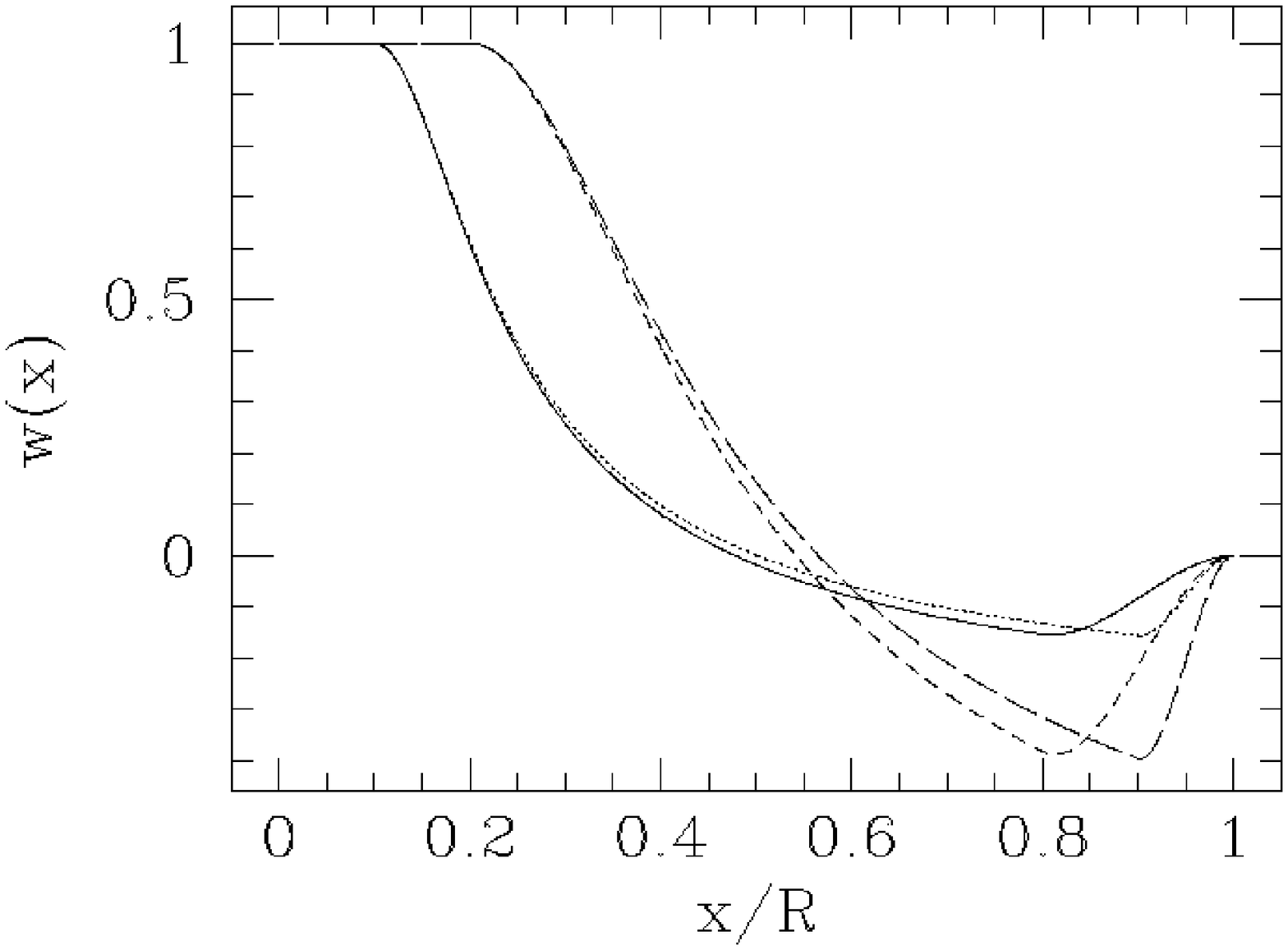}{8}
The corresponding function $Q(x)$ can be calculated from (16); again,
the resulting expressions are fairly cumbersome and shall not be
reproduced here. For the same values of $\nu_1$ and $\nu_2$ as in
Fig.\ts 1, the corresponding function $Q(x)$ is plotted in Fig.\ts 2.
\xfigure{2}{For the same values of $\nu_1$ and $\nu_2$ as in Fig.\ts
1, the corresponding function $Q(x)/x^2$ is plotted as a function of
$x/R$. The identification of the curves is the same as in Fig.\ts 1}
{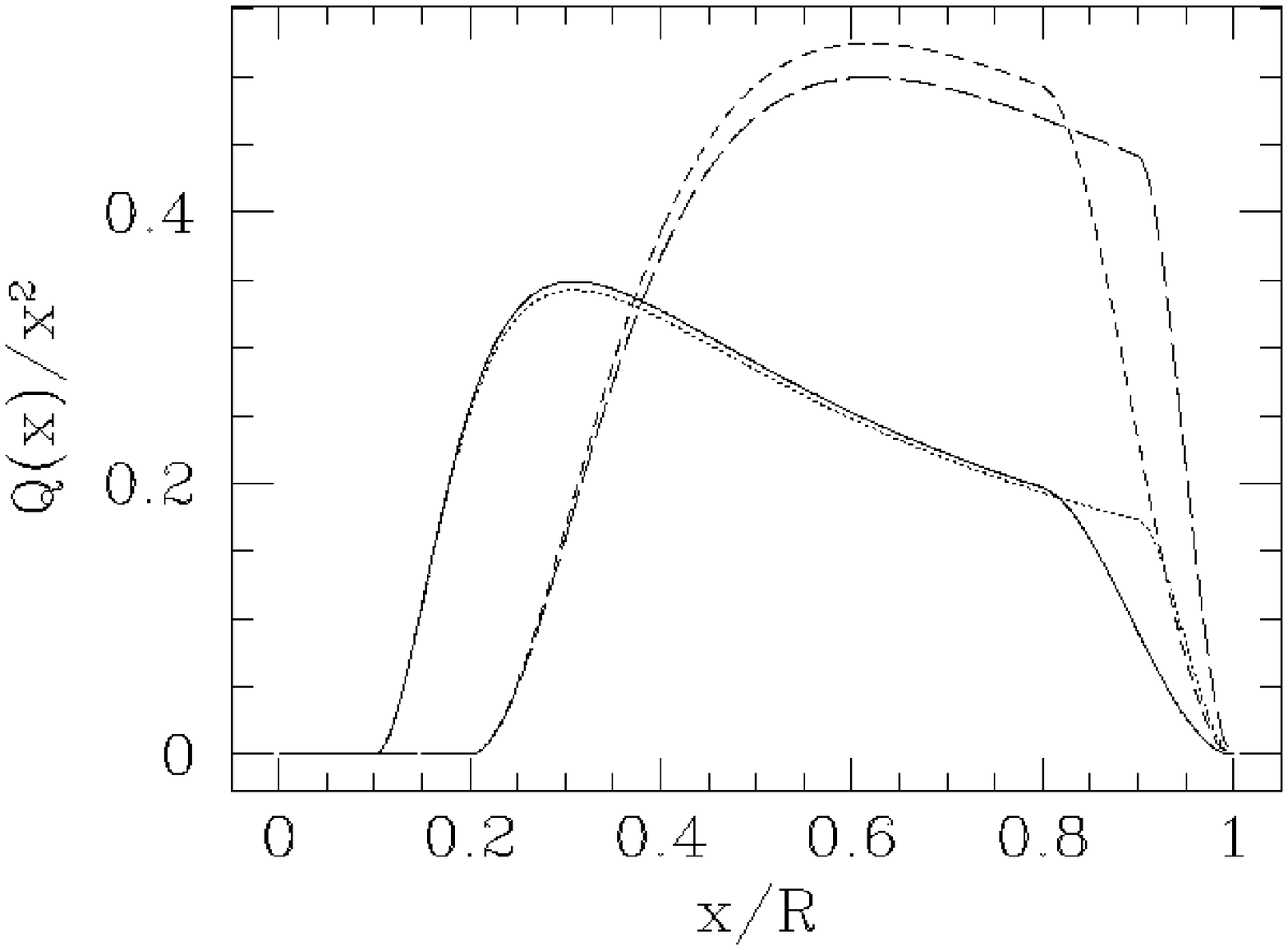}{8}
Finally, the special case of a singular isothermal sphere is
considered. In this case, $\ave{\gamma_{\rm t}}=x_0/(2 x)$, where
$$
x_0=4\pi\rund{\sigma_v\over c}^2 {D_{\rm ds}\over D_{\rm s}}
\eqno (35)
$$
is the Einstein radius of the mass distribution. Using this tangential
shear distribution in (28), one finds
$$
S_{\rm c}={\sqrt{\pi n x_0^2}\over \sigma_\eps}
{\int_0^R\d x\;Q(x)/x^2 \over
\sqrt{\int_0^R\d x\;Q^2(x)/x^3}}
=: s\,{\sqrt{\pi n x_0^2}\over \sigma_\eps}\quad .
\eqno (36a)
$$
Noting that the ratio of integrals does not depend on $R$ for the
function $Q(x)$ corresponding to $w(x)$ in (34), we have the
remarkable fact that the signal-to-noise for a singular isothermal
sphere is independent of the aperture size $R$, but depends only on
{\it the expected number of galaxies within the Einstein radius $x_0$
and the intrinsic ellipticity dispersion of the sources.} For the
weight function (18) defined by Kaiser, $Q=0$ for $x\le \nu_1 R=x_1$,
and $Q=$\ts const. for $x\in[\nu_1 R, R]$. Also in this case does the
ratio $s$ of integrals not depend on $R$, and one can easily calculate
that then $s=\sqrt{2(1-\nu_1)/(1+\nu_1)}$. The ratio $s$ is plotted in
Fig.\ts 3 as a function of $\nu_1$, for $\nu_2=0.9$; in fact, the
ratio $s$ depends only weakly on $\nu_2$, as long 
as $(1-\nu_2)\ll 1$. As can be seen, reducing $\nu_1$ increases the
signal-to-noise considerably for the weight function (34); for
$\nu_1\gtrsim 0.1$, the Kaiser weight function yields larger
signal-to-noise ratios. This is due to the fact that for larger values
of $\nu_1$, the weight function (34) deviates significantly from the
optimal weight function (32). Since the choice (34) of $w$ is by no
means unique, one might try to construct functions which yield a
higher signal-to-noise for ones favourite mass distribution. In this
paper, we shall use $w$ as given by (34), and restrict out attention
to small values of $\nu_1$. 
\xfigure{3}{The ratio $s$ as defined in (36a), as a function of
$\nu_1$. The solid curve is for the weight function $w(x)$ as defined
in (34), for $\nu_2=0.9$, whereas the dashed curve has been calculated
for the weight function (18) as defined by Kaiser (1995)}{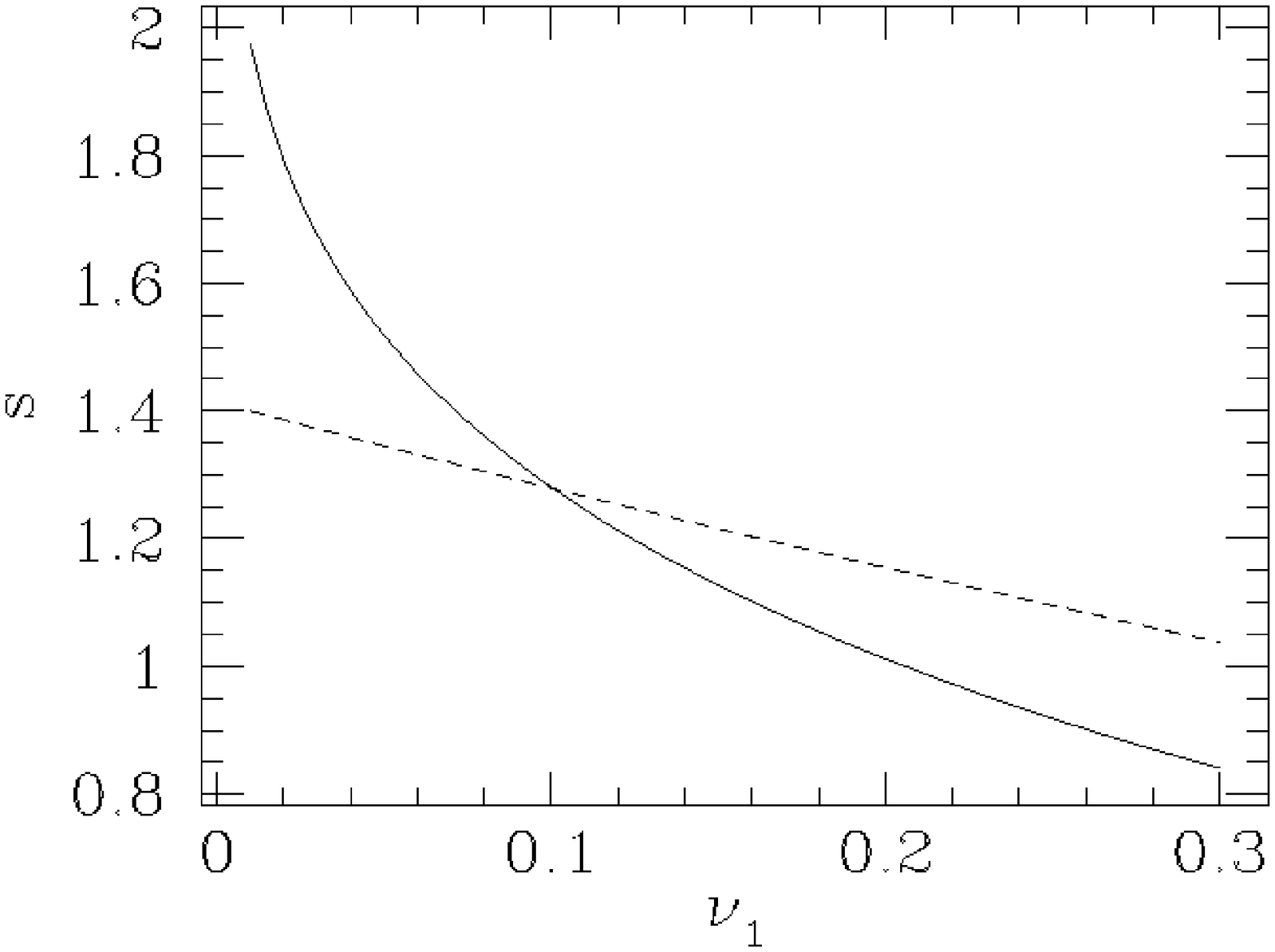}{8}
For reference, we rewrite (36a) in useful units as
$$
S_{\rm c}=8.39\, s \rund{n\over 30/\hbox{arcmin}^2}^{1/2}
\rund{\sigma_v\over 600\hbox{km/s}}^2 \rund{\sigma_\eps\over 0.2}^{-1}
\rund{D_{\rm ds}\over D_{\rm s}}\quad.
\eqno (36b)
$$

\sec{4 Detection of mass concentrations from synthetic data}
In this section the finding of mass concentrations from the shear
pattern in faint galaxy data is discussed. For this purpose, synthetic
data are generated, which consist of a set of position vectors $\vc
x_i$ of galaxy images and their complex ellipticity $\eps_i$. From
these data, a signal-to-noise map is obtained, by calculating the
function $S(\vc x)$ as defined in (24) on a grid. Mass concentrations
should show up as significant peaks in the $S$-map. 

Galaxies are distributed randomly on a square of side $x_{\rm f}$;
assuming a number density of $n$, $N=n x_{\rm f}^2$ galaxies are
placed on this square. For each such galaxy, the intrinsic ellipticity
$\eps^\s$ is drawn from the distribution function
$$
p_{\rm s}(\epsilon^\s)={1\over \pi 
\sigma_\eps^2(1-\Rm e^{-1/\sigma_\eps^2})}
\Rm e^{-\abs{\epsilon^\s}^2/\sigma_\eps^2}\quad,
\eqno (37)
$$
and the value $\sigma_\eps=0.2$ is taken throughout this paper (see
Seitz \& Schneider 1996, and references therein). For a given lens
model, the surface mass density $\kappa(\vc x_i)$ and the shear
$\gamma(\vc x_i)$ is calculated at each galaxy position, and the
`observed' ellipticity $\eps_i$ is calculated from (19). By selecting
the galaxy positions randomly on the observer's sky, we have implicity
assumed that the magnification bias can be neglected, i.e., that the
number counts of galaxies follows approximately the law $\d\log
N(m)/ \d m \propto 10^{\gamma m}$, with $\gamma\approx 0.4$. 
For a given filter size $R$, the $S$-statistics is calculated on a
square of size $x_{\rm f}-2 R$, to avoid boundary effects.

\subs{4.1 Lens-free $S$-statistics}
We first consider the case that no lenses are present, i.e, that
$\eps_i=\eps^\s_i$. In this case one expects that the $S$-statistics
(24) has a Gaussian probability distribution of unit width,
$$
p(S)={1\over \sqrt{2 \pi}}\Rm e^{-S^2/2}\quad .
\eqno (38)
$$
To verify this expectation, a simulation was carried out with $x_{\rm
f}=60\arcmin$, a number density of $n=30/{\rm arcmin}^2$ (this value
will be used throughout the rest of this paper), and a grid of
$1500^2$ points at which $S$ was calculated. The resulting probability
distribution $p(S)$ is plotted in Fig.\ts 4, together with the
Gaussian (38). As can be seen, the probability distribution of $S$
follows the expected Gaussian curve within the numerical errors.
\xfigure{4}{The probability distribution $p(S)$ is plotted (dashed
curve), together with the Gaussian distribution (38) (solid
curve). Also, the distribution function for the $S'$ statistics as
defined in the text is plotted (dotted curve), which shows somewhat
broader wings than the $S$-statistics. For these curves, $R=0\arcminf
5$ was used, with $\nu_1=0.1$ and $\nu_2=0.9$}{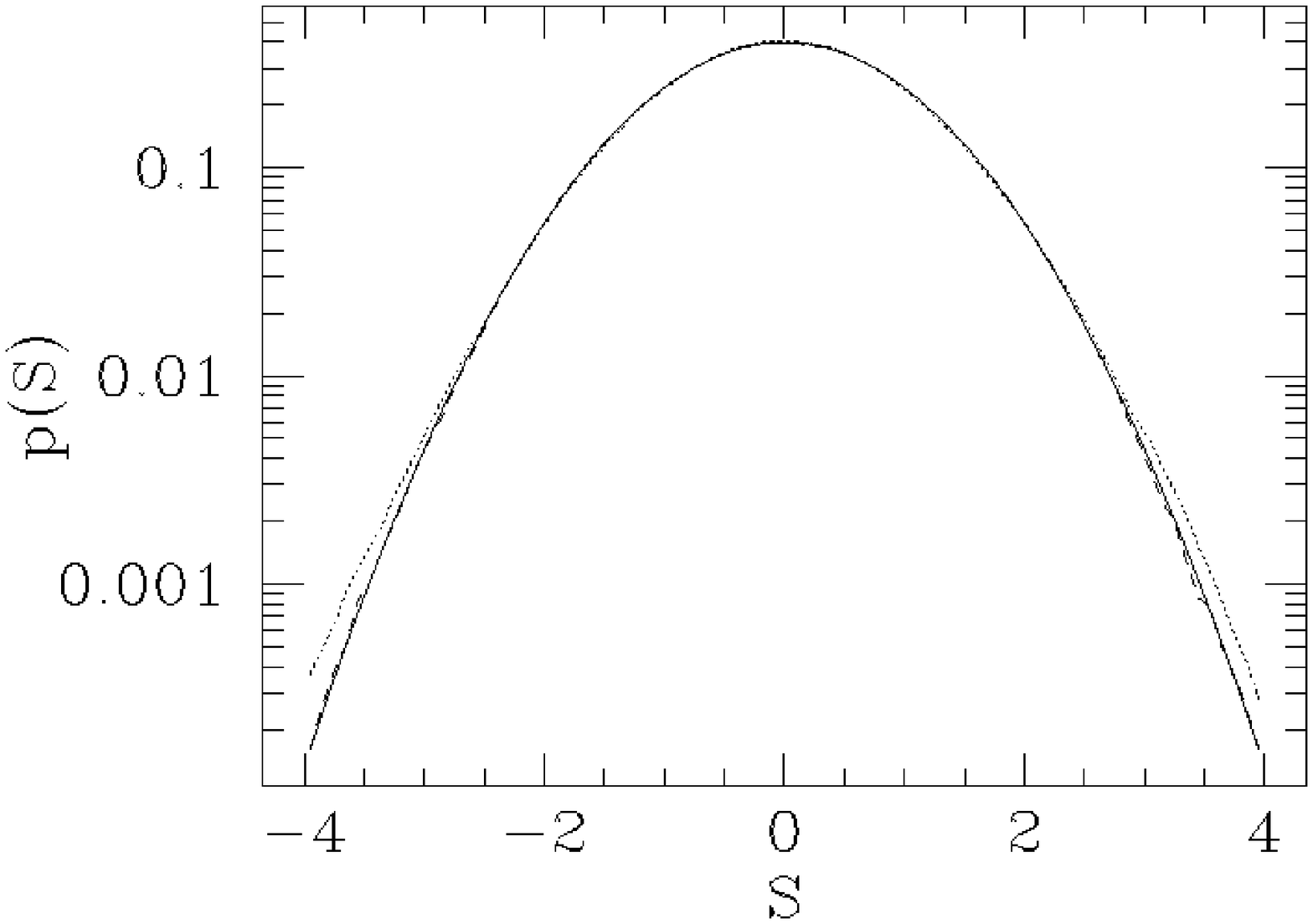}{8}
In addition, a modified $S$-statistics was calculated
which uses $S'=m/\sigma_{m{\rm c}}$ instead of $S=m/\sigma_{m{\rm
d}}$. Whereas the former is slightly more faster to evaluate, since
$\sigma_{m{\rm c}}$ is independent of $\vc x_0$, Fig.\ts 4 shows
that the probability distribution for $S'$ has somewhat broader wings
than that of $S$. The same simulations have been carried out with the
weight function $w_{\rm K}$, and the results are basically
indistinguishable from those shown in Fig.\ts 4. 
\xfigure{5}{Contour plots of the signal-to-noise 
ratio $S$ defined in Eq.\ts(24),
for a surface number density of $n=30/{\rm arcmin}^2$, on a square of
sidelength $10\arcmin$. No lensing has been assumed, i.e., the galaxy
images are distributed randomly with an ellipticity distribution
according to (37). In the upper left panel, $R=0\arcminf 5$, in
the upper right panel, $R=1\arcmin$, and in the lower left panel,
$R=2\arcmin$. In these three cases, the weight function (34) was used,
with $\nu_1=0.1$, $\nu_2=0.9$. The lower right panel uses the weight
function $w_{\rm K}$ given in (18), for $R=1\arcmin$ and
$\nu_1=x_1/R=0.3$. Contour levels are $S=1,2,3,\dots$ (solid
contours) and $S=-1,-2,-3,\dots$ (dashed contours)}{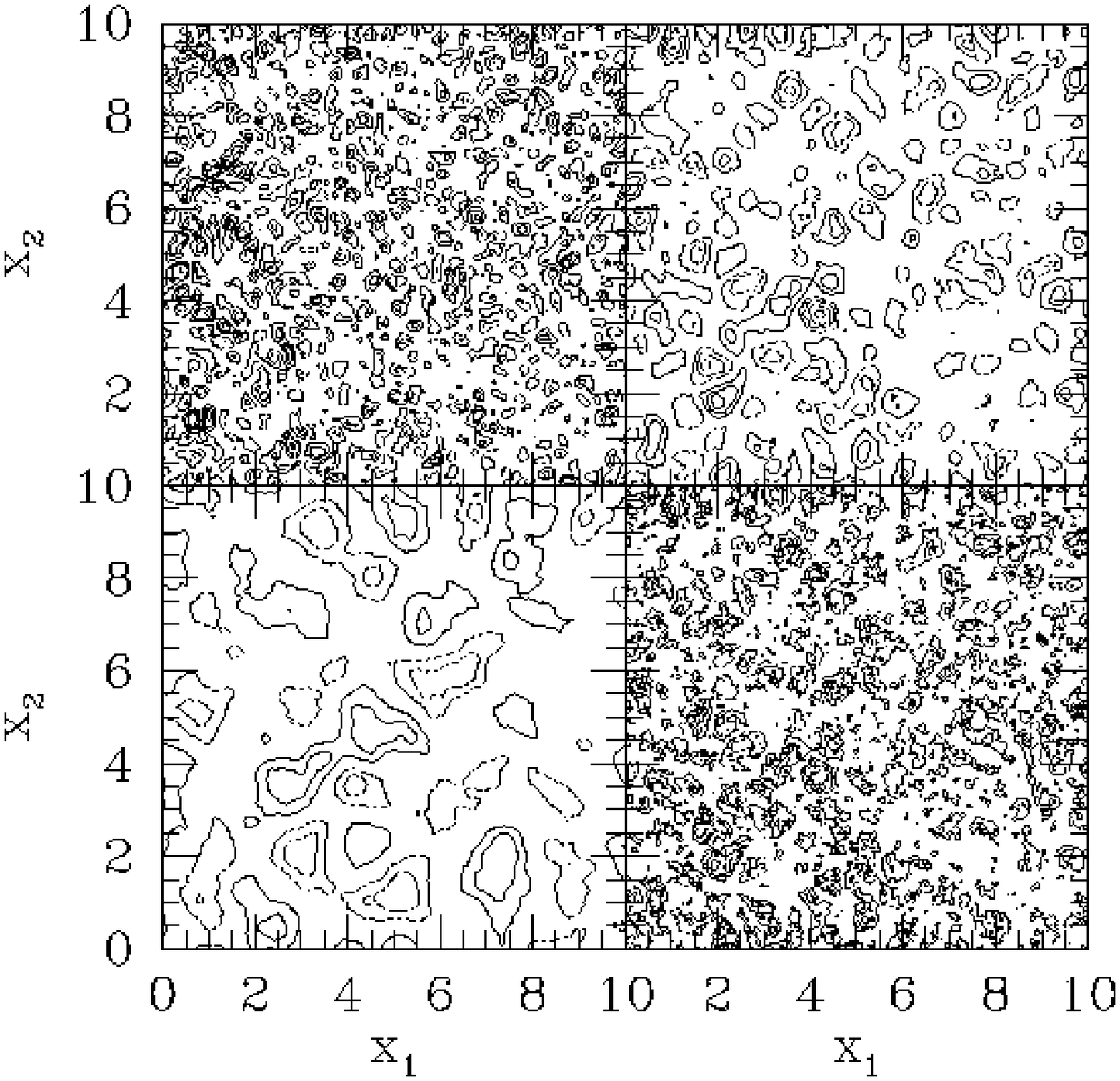}{16}
To illustrate the spatial distribution of the function $S$, Fig.\ts 5
displays contour maps of $S$ on a square of sidelength $10\arcmin$;
i.e., we have chosen $x_{\rm f}=10\arcmin +2R$. No lensing has been
assumed. Three different filter scales $R$ are chosen in Fig.\ts5 for
the weight function (34); in addition, an $S$-map for the filter
$w_{\rm K}$ given in (18) has been plotted in the lower right
panel. Obviously, by increasing $R$, the $S$-maps become smoother. It
should be noted that the $S$-map for the $w_{\rm K}$ weight function
is much `noisier' than the one for the $w$ weight
function for the same filter size $R$ (the two right-hand panels in
Fig.\ts 5). This is because the weight function $w_{\rm K}$, and thus
the corresponding function $Q$, are discontinuous, so that single
galaxy images moving into the filter can cause sudden jumps of $S$.
In the maps with small $R$, several peaks with $S>3$ (or
minima with $S<-3$) can be seen, but extrema with $\abs{S}>4$ are very
rare, in accord with the result shown in Fig.\ts 4. 

\subs{4.2 Single-lens statistics}
Next we study the statistical properties of the signal-to-noise $S$
for a single isolated deflector. For this, a mass model is chosen
which closely resembles an isothermal sphere with a finite core and a
finite truncation radius, described by the (axially-symmetric) mass
distribution 
$$
\kappa(x)={x_0\over 2\sqrt{x^2+x_{\rm c}^2}}-{x_0\over
2\sqrt{x^2+x_{\rm t}^2}}\quad, 
\eqno (39)
$$
where $x_{\rm c}$ is the core radius, $x_{\rm t}$ the truncation
radius, and $x_0$ is the Einstein angle of the corresponding singular
isothermal model, given in (35). For the rest of the paper, a fixed
value of $D_{\rm ds}/D_{\rm s}=0.7$ is assumed. The (tangential) shear
corresponding to this mass distribution is calculated as
$\gamma\equiv\gamma_{\rm t}=\bar\kappa -\kappa$, where 
$$
\bar\kappa(x)={x_0\over x^2}\rund{\sqrt{x^2+x_{\rm c}^2}+x_{\rm
t}-\sqrt{x^2+x_{\rm t}^2} -x_{\rm c}}\quad .
$$
For this mass model, the expected signal-to-noise ratio $S_{\rm c}$
can be calculated from (28); for different values of $\sigma_v$ and
the ratio $x_{\rm c}/x_0$, this expected value is plotted as a
function of the filter scale $R$ in Fig.\ts 6. In addition,
simulations were made by distributing galaxies around these lenses
with an intrinsic ellipticity distribution (37), then distorting them
by the lens using (19), and calculating $S$ according to (24) at the
center of the lens, again for varying filter scale $R$. For different
realizations of the galaxies, the resulting values of $S$ are also
plotted in Fig.\ts 6.
\xfigure{6}{The values of $S$ -- see Eq.\ts (24) -- for different
realizations of the galaxy distribution and different filter radii
$R$. The numbers in each panel shows the pair of values
$(\sigma_v,x_{\rm c}/x_0)$ (with $\sigma_v$ in km/s)
used in the lens model (39). The truncation
radius is chosen to be $x_{\rm t}=50 x_0$. For the weight function
$Q$, the values $\nu_1=0.1$ and $\nu_2=0.9$ were chosen. The solid
curve in each panel is the ensemble-averaged signal-to-noise $S_{\rm
c}$ as given in (28). The systematic differences between the
numerical results (crosses) and the analytic expectation are due to
the non-linearity of the image distortion -- see text}{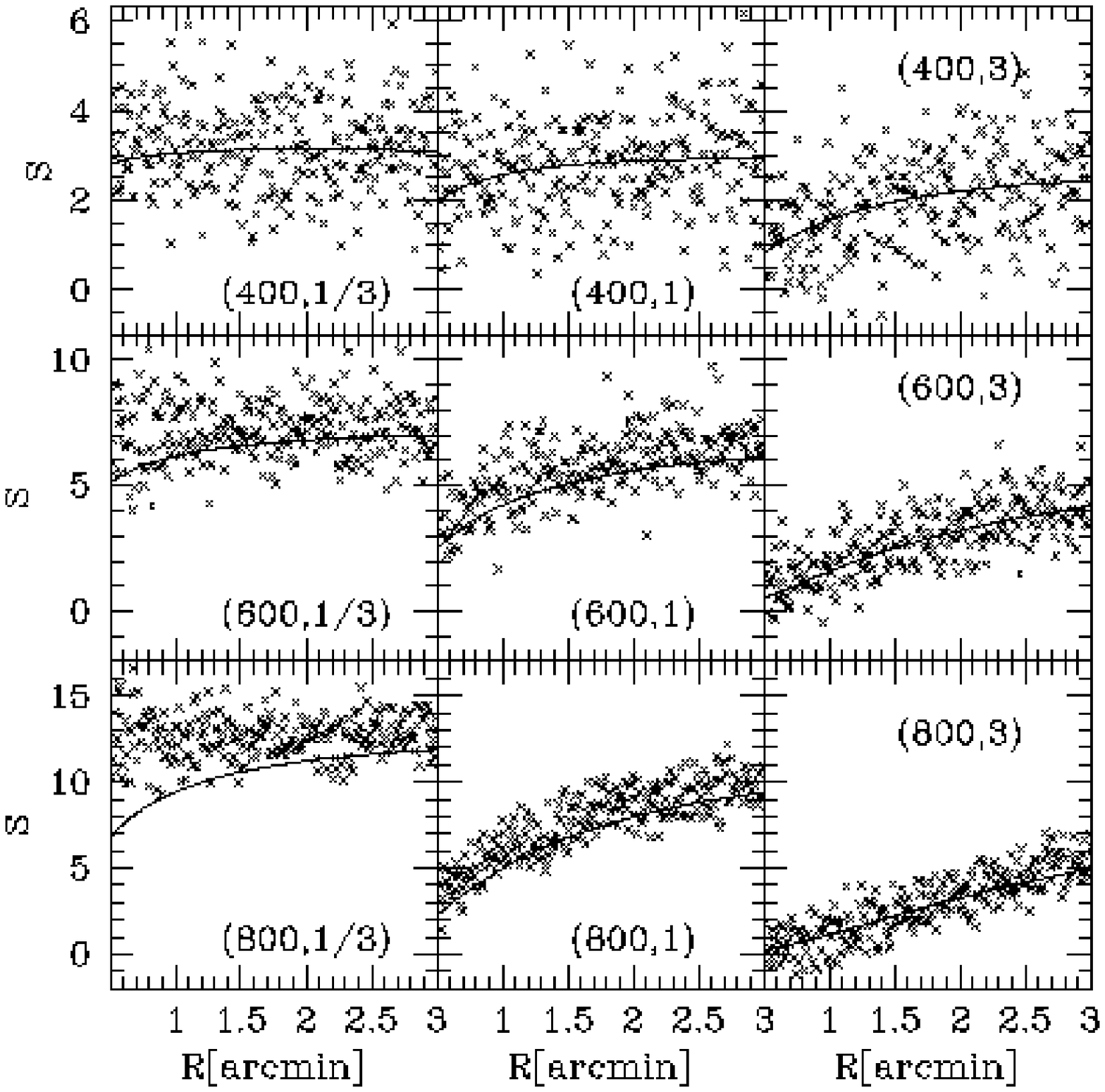}{15}
Three different cases are shown in Fig.\ts 6: for $x_{\rm c}/x_0=1/3$,
the lens is critical; it is nearly critical for $x_{\rm c}/x_0=1$, and
a rather weak lens for $x_{\rm c}/x_0=3$. 
The ensemble-averaged signal-to-noise $S_{\rm c}$ as calculated from
(28) lies well in the `middle' of the numerically-generated points for
the non-critical lens, but significantly underestimates $S$ for the
critical lenses. This is because the relation $g\approx\gamma$ -- see
(20) -- is no longer valid near the center of a critical lens.
In the calculation of (28) the true value of $\gamma$ is used, whereas
for calculating $S$ according to (24), $\eps_{\rm t}$ is
used. Now, $\eps$ is an unbiased estimate for $g$, which is larger in
magnitude that $\gamma$, so that $S_{\rm c}$ is biased toward lower
values for strong or even critical lenses. 

For critical 
lenses with $\sigma_v=400$\ts km/s, the mean signal-to-noise is of
order 3, as also follows from (36b). Hence, these lenses can be
detected with a marginal significance, given the fact that $S$ varies
considerably around this expectation value for individual
realizations. However, if the core radius becomes larger than about
$x_0$, these lenses are too weak to be reliably 
discovered individually by the method proposed here. Lenses with
$\sigma_v\gtrsim 600$\ts km/s can be discovered reliably if the core
size is not too large. In addition to the core radius, the filter
scale $R$ plays an important role: if this filter scale is not
considerably larger than the core radius, the $S$-statistics becomes
very insensitive, since it measures radial gradients. Hence, if the
filter has the same size as the core, no gradient can be measured. For
critical lenses, even the smallest filter scale considered in Fig.\ts
6 is sufficient, whereas for weak lenses with $x_{\rm c}=3 x_0$, the
filter scale $R$ must be larger than $2\arcminf 5$ to detect reliably
a lens with $\sigma_v=800$\ts km/s. For such large core size, lenses
with $\sigma_v=600$\ts km/s are not reliably detected, but those with
$x_{\rm c}=x_0$ are if $R\gtrsim 1\arcminf 5$. Note that the width of
the distribution of points in all panels of Fig.\ts 6 is of order
unity. 

\subs{4.3 $S$-statistics for an ensemble of lenses}
Next we want to investigate whether the $S$-statistics is
significantly degraded by not considering isolated lenses, but by
studying an ensemble of lenses. Two kinds of effects may affect the
$S$-statistics: small-scale mass concentrations which can locally
dominate the shear field and which should increase the noise in the
$S$-statistics, and larger-scale mass perturbations. The latter should
not provide a problem for the detection of mass concentrations as long
as their contribution can be described as a linear function across the
size of the filter (i.e., where the weight function $Q(x)$ does not
vanish). In order to study these effect, we have made simulations with
rather extreme perturbations. Motivated by Fig.\ts 6, it should be
investigated whether lenses with the density profile (39) and
with $\sigma_v=600$\ts km/s and $x_{\rm
c}=x_0$ can be significantly detected; those will be called `main
lenses' in the following. They were distributed randomly with a number
density of 16/deg$^2$. In addition, small-scale lenses were added with
$\sigma_v=220$\ts km/s and $x_{\rm c}=x_0$, and a density which is 100
times higher than that of the main lenses. To simulate larger-scale
perturbations, lenses with $\sigma_v=800$\ts km/s and $x_{\rm c}=5
x_0$ were distributed with twice the density of the main
lenses. Judging from Fig.\ts 6, these `perturbing' mass distributions
should not be reliably identifyable in the $S$-map.
\xfigure{7}{A map of the $S$-statistics, with filter scale
$R=2\arcmin$, for the lens distribution as
described in the text. Contours are $S=1,2,3,\dots$, crosses denote
the positions of the `main' lenses with $\sigma_v=600$\ts km/s and
$x_{\rm c}=x_0$, and triangles the position of the
`perturbing' lenses with $\sigma_v=800$\ts km/s and $x_{\rm c}=5 x_0$}
{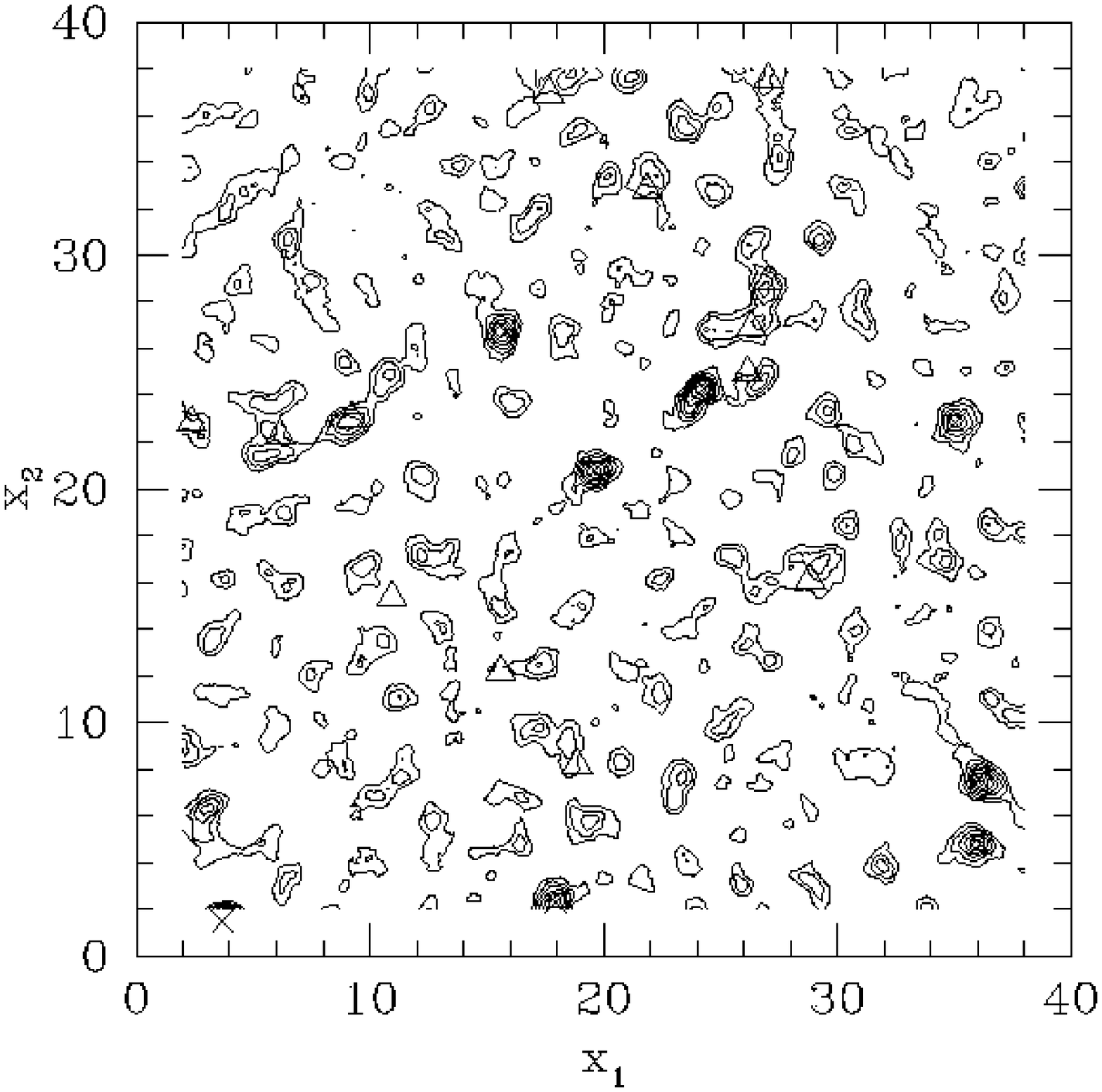}{16}
As can be seen from Fig.\ts 7, several peaks occur with $S\gtrsim 5$,
and all of them correspond to one of the `main lenses'; in fact, there
is no single peak with $S\ge 5$ which is not located at a `main lens'
position. Some of the large-scale perturbing mass distributions,
indicated by triangles in Fig.\ts 7, produce local peaks with
$3\lesssim S\lesssim 5$, but these peaks cannot be confused with those
corresponding to the `main deflectors'. From simulations of a larger
angular field, the statistics of peaks -- as shown in Fig.\ts 8 --
clearly indicates that all `main lenses' are identified as $S\ge 5$
peaks, and none of the peaks with $S\ge 5$ is a `false detection'. We
can thus conclude that the aperture method for the detection of matter
concentrations does work not only for isolated axially-symmetric mass
distributions, but also if one accounts for large- and small-scale
`perturbations' of the lens mapping.
\xfigure{8}{Peak statistics: on a field 
with side $x_{\rm f}=80\arcmin$, lenses with $\sigma_v=600$\ts km/s
and $x_{\rm c}=x_0$ (``main lenses'') were distributed with a density
of 16/deg$^2$; in addition, to simulate small-scale noise, hundred times
as many lenses with $\sigma_v=220$\ts km/s and $x_{\rm c}=x_0$, and
for simulating large-scale perturbations, two times as many lenses
with $\sigma_v=800$\ts km/s and $x_c=5 x_0$ were distributed. On the
resulting $S$-map, all peaks with $S\ge 3$ were identified, and their
separation $D_{\rm min}$ to the nearest main lens was
calculated. Plotted are the peaks of the $S$-map in the $D_{\rm
min}$-$S_{\rm peak}$ plane. Circles corresponds to those peaks which
are the closest ones to a main lens, and crosses denote the
others. Here, a filter scale of $R=2\arcmin$ was used. All peaks with
$S_{\rm peak}\ge 5$ are situated within $R/7$ from a
main lens; hence, all peaks with
$S_{\rm peak}\gtrsim 5$ correspond to a mass concentration. Of the 29
main lenses in the field, the lowest one has $S_{\rm peak}=5.36$, 
three have $S_{\rm peak}\in[5,6]$, 
nine have
$S_{\rm peak}\in [6,7]$, 13 have $S_{\rm peak}\in [7,8]$, and the
other four main lenses have $S_{\rm peak}\ge 8$. Thus, not only are
there no false detections of main lenses for $S_{\rm peak}\ge 5$, but
also all the main lenses are detected with $S_{\rm peak}\ge 5$}
{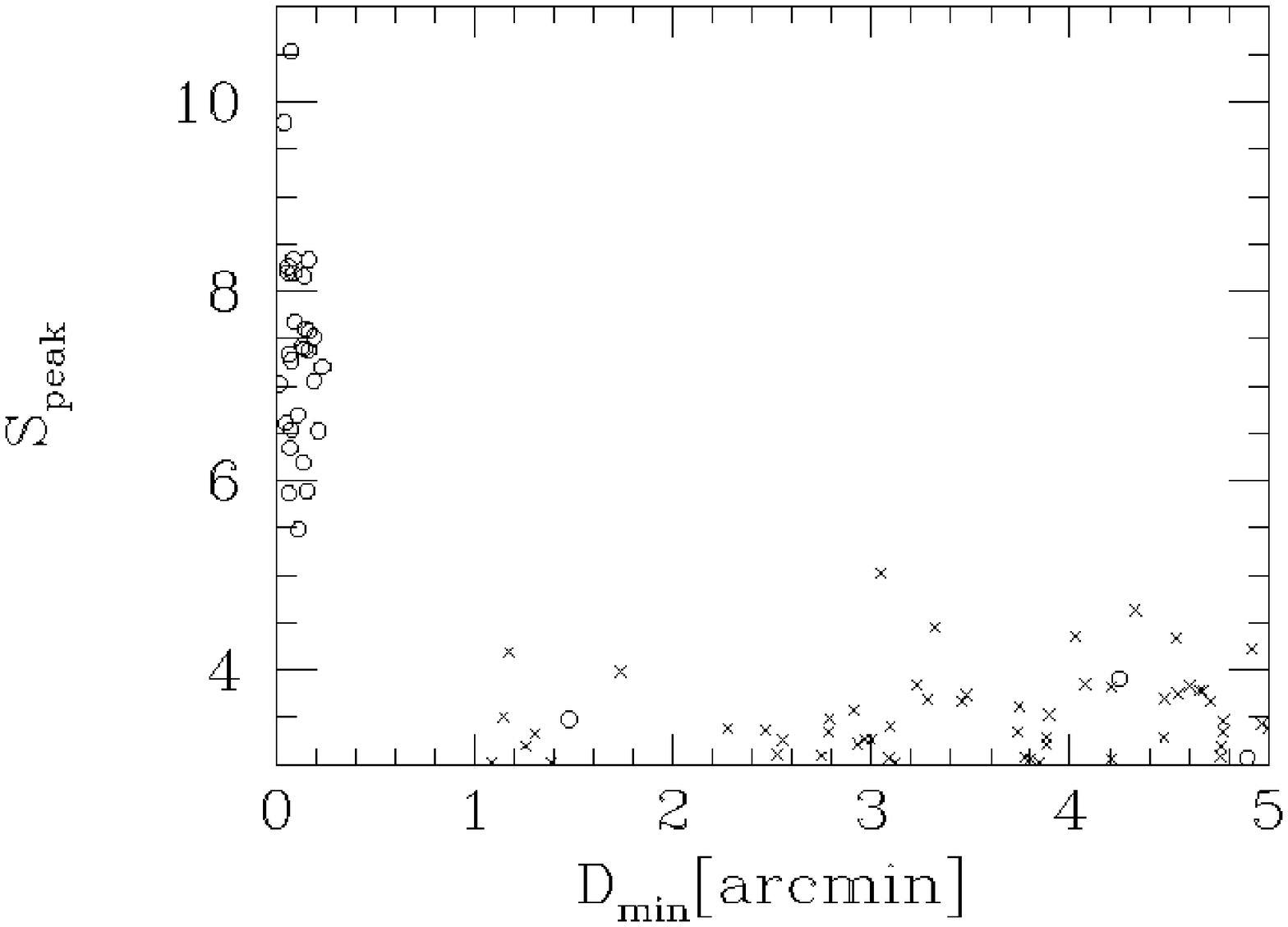}{11}

\def\o{{(\rm obs)}}
\def\psf{{(\rm PSF)}}

\subs{4.4 Effects of a point-spread function}
One of the most difficult problems for weak lensing studies from
ground-based observations is the proper accounting for a point-spread
function (PSF). The observed image ellipticity $\eps^\o$ deviates from
the true image ellipticity $\eps$, depending on the size of the PSF
and the size of the image. Small elliptical images are transformed
into nearly round observed images by an isotropic PSF. In addition,
anisotropies of the PSF can introduce spurious shear signals in the
data. For these reasons, cluster-mass reconstructions from
ground-based images have to correct for PSF effects (e.g., Bonnet \&
Mellier 1995; Kaiser, Squires \& Broadhurst 1995). Depending on the
seeing conditions, rather large correction factors have to be used to
translate the `observed shear' to the `true shear'. In this
subsection, it will be demonstrated that PSF effects are much less
important for the detection of dark matter concentrations if the
method described before is applied.

In order to investigate the effects of a PSF on the observed image
ellipticity, we note that the complex ellipticity is defined in terms
of the second moment tensor $Q_{ij}$ of the image brightness
distribution as
$$
\eps={Q_{11}-Q_{22}+2\Rm i Q_{12} \over 
Q_{11}+Q_{22}+2\sqrt{Q_{11}Q_{22}-Q_{12}^2}}\quad,
\eqno (40)
$$
where 
$$
Q_{ij}={\int \d^2 x\; I(\vc x)\,(x_i-\bar x_i)(x_j-\bar x_j) \over
\int \d^2 x\; I(\vc x)}\quad,
\eqno (41a)
$$
and $\vc {\bar x}$ is the `center' of the image,
$$
\vc{\bar x}={\int \d^2 x\; I(\vc x)\,\vc x\over \int \d^2 x\; I(\vc
x)}\quad.
\eqno (41b)
$$
In a similar way, we define the complex ellipticity $\eps^\o$ in terms
of the second moment tensor $Q_{ij}^\o$, which is defined in terms of
the observed brightness profile
$$
I^\o(\vc x)=\int\d^2 y\;I(\vc y)\,P(\vc x-\vc y)\quad,
\eqno (42)
$$
where $P(\vc x)$ is the PSF, which is assumed to be normalized. From
these relations, one finds that the relation between $Q_{ij}$ and
$Q_{ij}^\o$ reads
$$
Q^\o_{ij}=Q_{ij}+P_{ij}\quad ,
\eqno (43)
$$
where 
$$
P_{ij}=\int \d^2 z\;P(\vc z)\,(z_i-\bar z_i)(z_j-\bar z_j)\quad ,
\eqno (44a)
$$
and
$$
\vc{\bar z}=\int\d^2 z\;P(\vc z)\,\vc z\quad .
\eqno (44b)
$$
From these relations, one can obtain $\eps^\o$ in dependence of
$\eps$. A slightly more convenient way to find this relation is to
consider the ellipticity variable
$$
\chi={Q_{11}-Q_{22}+2\Rm i Q_{12} \over 
Q_{11}+Q_{22}}={2\eps\over 1+\abs{\eps}^2}\quad,
\eqno (45)
$$
and correspondingly for $\chi^\o$. Then, in agreement with a similar
result derived in Bartelmann (1995) and Villumsen (1995),
$$
\chi^\o={\chi+T \chi^\psf \over 1+T}\quad,
\eqno (46)
$$
where $T$ describes the ratio of the size of the PSF and the size of
the image, and $\chi^\psf$ accounts for a possible anisotropy of the
PSF, 
$$
T={P_{11}+P_{22}\over Q_{11}+Q_{22}}\; ;\;
\chi^\psf={P_{11}-P_{22}+2\Rm i P_{12}\over P_{11}+P_{22}}\quad.
\eqno (47)
$$
Hence, (46) expresses the fact that for small $T$ (i.e., size of the
PSF small compared to the image size), the observed ellipticity is
approximately equal to the true ellipticity of the image, whereas for
large $T$, the observed ellipticity can be dominated by the
ellipticity of the PSF. Using the inverse of (45),
$$
\eps={\chi\over 1+\sqrt{1-\abs{\chi}^2}}\quad,
\eqno (48)
$$
one can now calculate $\eps^\o$ as a function of $\eps$, for a given
PSF and for a given image size. We define the size $\theta$ of an
image, and the size $\sigma$ of the seeing disk as
$$
\theta:=\sqrt{Q_{11}+Q_{22}}\quad ; \quad
\sigma^2=P_{11}+P_{22}\quad,
\eqno (49)
$$
so that $T=\sigma^2/\theta^2$. 

As in the idealized case of negligable seeing, we define in analogy to
(22)
$$
m^\o(\vc x_0)={1\over n}\sum_i {\eps_{{\rm t}i}^\o(\vc x_0)\over
\abs{\vc x_i-\vc x_0}^2}\,Q\rund{\abs{\vc x_i-\vc x_0}}\quad .
\eqno (50)
$$
Whereas $m^\o$ will deviate significantly from $m$ due to seeing
effects, and thus no longer is an approximation for the $w$-weighted
mass density as defined in (9), one can still use $m^\o$ to measure
local mass overdensities; in particular, one can in analogy to (24)
define the signal-to-noise ratio
$$
S^\o(\vc x_0)={\sqrt{2}\over \sigma_\eps^\o}
{\sum_i {\eps_{{\rm t}i}^\o(\vc x_0)\over
\abs{\vc x_i-\vc x_0}^2}\,Q\rund{\abs{\vc x_i-\vc x_0}} \over
\sqrt{ \sum_i {Q^2\rund{\abs{\vc x_i-\vc x_0}}\over 
\abs{\vc x_i-\vc x_0}^4}}} \quad,
\eqno (51)
$$
where $\sigma_\eps^\o$ is the dispersion of the {\it observed}
ellipticities. Hence, whereas the `signal' $m^\o$ will be smaller due
to seeing than $m$, so will be the dispersion, and one might expect
that the signal-to-noise ratio is not as badly affected as the signal
itself. 
\xfigure{9}{The signal-to-noise ratio $S^\o$ as a function of the
ratio of the size $\sigma$ of the seeing disk and the characteristic
size $\theta_0$ of the galaxy distribution. Here, the size
distribution (54a) was assumed. An `isothermal' lens model of the form
(39) was assumed, with $x_{\rm c}=x_0$ and $x_{\rm t}=50 x_0$. In the
upper panels, $\sigma_v=600$\ts km/s, in the lower panels,
$\sigma_v=800$\ts km/s. The filter size $R$ in the two left panels is
$R=1\arcminf 5$, and $R=3\arcmin$ in the right panels. The value of
$S^\o$ is plotted as crosses for an isotropic PSF (i.e.,
$\chi^\psf=0$), whereas the circles show the values for a PSF with
$\chi^\psf=0.1$. As in the rest of this paper, the number density of
galaxy images was taken to be 30/sq.\ts arcmin.
The solid curve displays the ensemble-averaged
signal-to-noise ratio, $\ave{m^\o}/\sigma_{m{\rm c}}$, calculated for
an isotropic PSF}{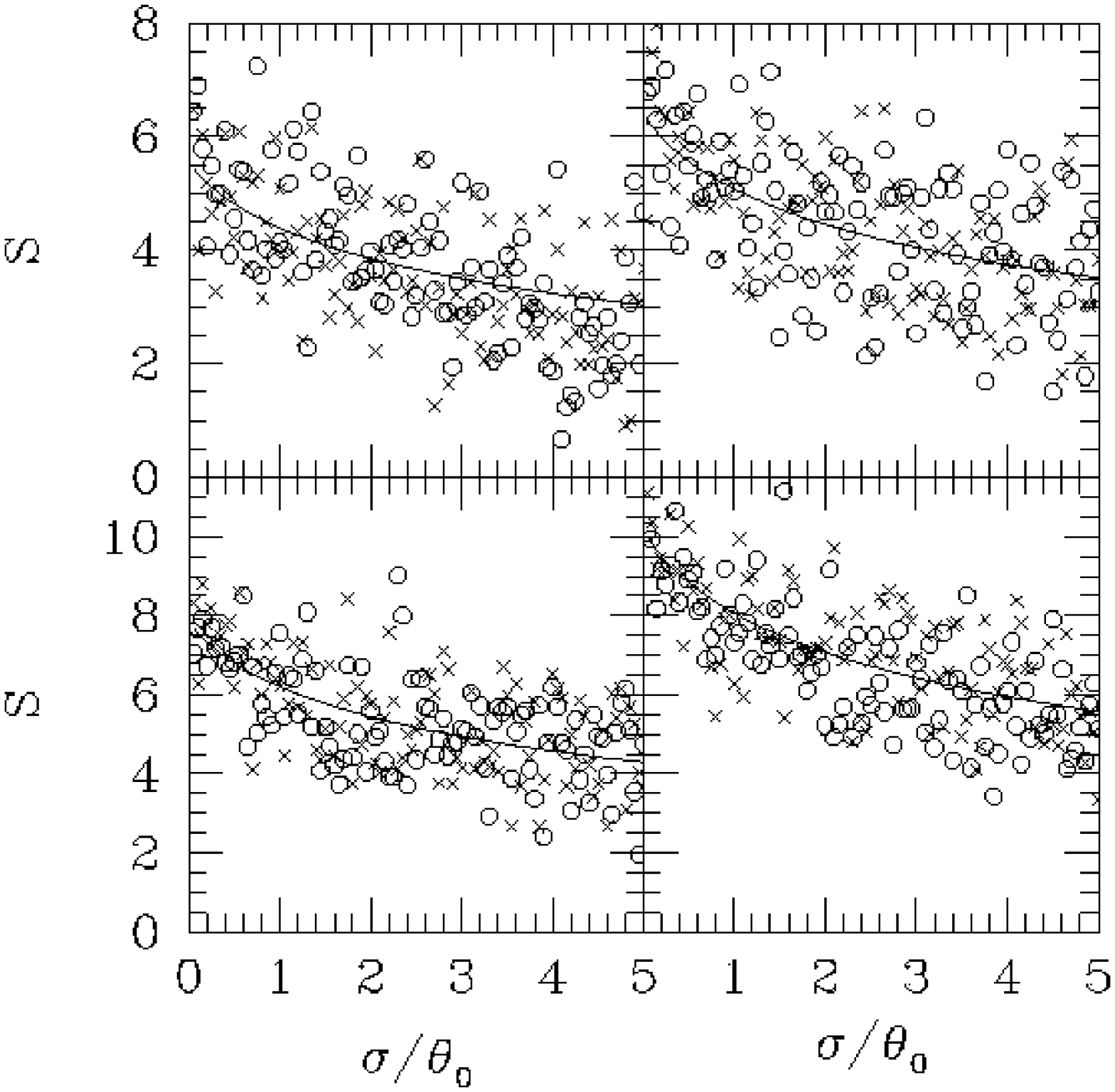}{14}
For an assumed size distribution $p_\theta(\theta)$ of the images, one
can define the expectation value of $m^\o$ for the center of an
axially-symmetric mass distribution 
(again, we shall drop the
argument $\vc x_0$ in the following) as
$$\eqalign{
\ave{m^\o}&={1\over n}\eck{\prod_{k=1}^N\int\d\theta_k\;p_\theta(\theta_k)
\int \d^2 \eps^\s_k\;p_{\rm s}\rund{\eps^\s_k}
\int{\d^2 x_k\over \pi R^2}}m^\o \cr
&=2\pi\int\d\theta\;p_\theta(\theta)
\int \d^2 \eps^\s\;p_{\rm s}\rund{\eps^\s}
\int_0^R \d x\;{\eps_{\rm t}^\o\,Q(x) \over x}\quad .\cr }
\eqno (52)
$$
Note that in this equation, $\eps_{\rm t}^\o$ is a function of
$\eps^\s$ and the reduced shear $g(x)$ through (19), i.e., to
determine the true image ellipticity $\eps$, and of $\theta$ (and the
PSF) through the transformation (45), (46) and (48) from $\eps$ to
$\eps^\o$. Because of the delicate dependencies, no further analytic
progress can be made, and (52) has to be evaluated
numerically. Similarly, one can calculate the dispersion of $m^\o$ in the
absence of a lens,
$$
\sigma^2_{m{\rm c}}={2\pi\over n}\int\d\theta\;p_\theta(\theta)
\int \d^2 \eps^\s\;p_{\rm s}\rund{\eps^\s}\,\rund{\eps_{\rm t}^\o}^2
\int_0^R {\d x\;Q^2(x)\over x^3}\quad .
\eqno (53)
$$
\xfigure{10}{Same as Fig.\ts 9, except that now the size distribution
(54b) instead of (54a) was used}{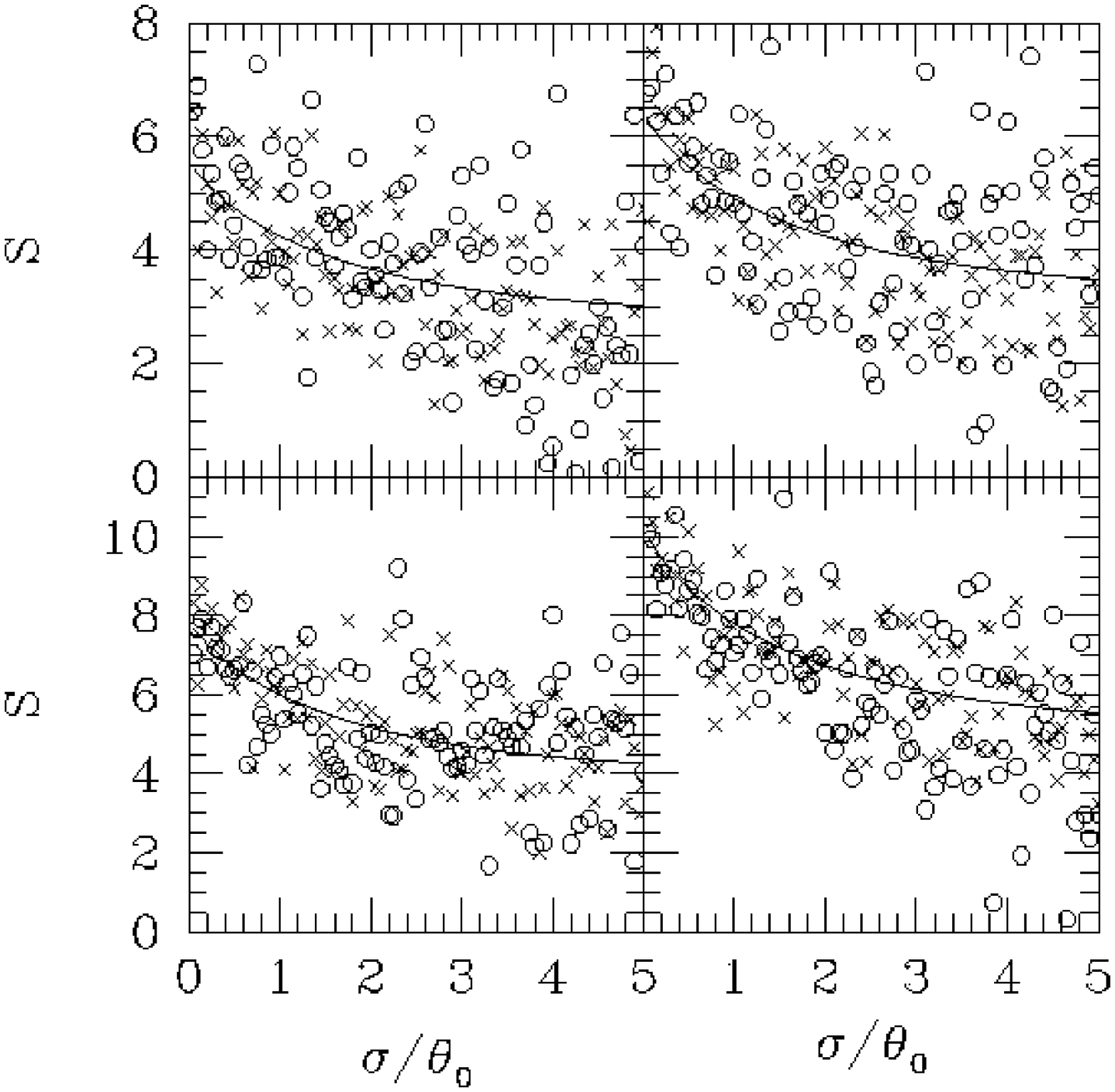}{14}
For the lens model described by (39), we have calculated the
ensemble-averaged signal-to-noise, $S_{\rm
c}=\ave{m^\o}/\sigma_{m{\rm c}}$, according to (52) and (53) for an
isotropic PSF,
and $S^\o$ for different realizations of the galaxy distribution, as a
function of the ratio $\sigma/\theta_0$, where $\theta_0$
characterizes the size distribution of the galaxy images. In
particular, we have chosen the following two distributions:
$$
p_\theta(\theta)={\theta_0\over \rund{\theta+\theta_0}^2}\quad,
\eqno (54a)
$$
or
$$
p_\theta(\theta)={1\over \theta_0}\Rm e^{-\theta/\theta_0}\quad .
\eqno (54b)
$$
It is obvious that $S$ depends only on the ratio $\sigma/\theta_0$,
since there are no other relevant scales in the problem. The results
of these calculations are shown in Fig.\ts 9 for the distribution
function (54a), and in Fig.\ts 10 for the distribution function (54b). 

The symbols in Figs.\ts 9 and 10 show $S^\o$ for different
realizations of the background galaxy population, calculated from
(51), whereas the solid curve shows $S_{\rm c}=\ave{m^\o}/\sigma_{m{\rm c}}$,
for two values of the velocity dispersion $\sigma_v$ and two values
for the filter size $R$. Crosses denote values of $S^\o$ calculated
for an isotropic PSF, whereas the circles assume that the PSF is
anisotropic, characterized by $\chi^\psf=0.1$. As can be seen, most
clearly from Fig.\ts 10, the distribution of the circles is broader
relative to that of the crosses;
this is due to the fact that for an anisotropic PSF, the contribution
of a galaxy image to the sum in the 
numerator of (51) does not only depend on the
radial separation of the image from the point of interest, but also on
its polar coordinate, since $\chi^\psf$ enters $\eps_{\rm t}$ as a
term $\propto \cos(2\vp)$. Hence, the polar coordinates of the galaxy
images provides an additional source of `Poisson noise' in $S^\o$.
The effect of a PSF is stronger for the size
distribution (54b) than for (54a), for fixed $\theta_0$, which is due
to the fact that for (54a) the number of `large' sources is
considerably larger than for (54b). As can be inferred from the
figures, lenses with $\sigma_v=600$\ts km/s can still be reliably
detected if the seeing is not much larger than the characteristic size
$\theta_0$ of the galaxy distribution. However, even for
$\sigma\lesssim 5 \theta_0$ can lenses with $\sigma_v=800$\ts km/s be
discovered, independent of whether the PSF is isotropic or not. Thus
we see that the effect of a PSF is much less dramatic than might have
been expected. This is mainly due to the fact that not only the signal
is weakened by a PSF, but also the noise, so that the signal-to-noise
ratio is much more weakly affected by a PSF than the signal alone.

\sec{5 Discussion}
The generalization (9) and (15) of the aperture densitometry (1), first
derived  Kaiser et al.\ts (1994), no longer has
such an intuitive meaning as (1), but this generalization allows the
definition of a measure which yields the highest signal-to-noise for
the detection of mass concentrations with an assumed radial
profile. In addition, by choosing $w$ appropriately, one can construct
a smooth signal-to-noise map from an observed image. The weight
function (34) used throughout this paper was constructed by assuming
the dark matter concentrations to have approximately an isothermal
profile. Numerical simulations have then shown that dark matter halos
can be detected reliably provided their typical velocity dispersion
exceeds about 600\ts km/s (for an assumed density of galaxy images of
30/arcmin$^2$, typical of current deep images from 4-m class telescopes);
inclusion of seeing slightly weakens the
lensing signal, and, depending on the ratio of the size of the seeing
disk and the characteristic image size of faint background galaxies,
slightly more massive mass concentrations are necessary for a robust
detection. Even a significant anisotropy of the PSF does not seriously
affect the signal-to-noise ratio as long as the PSF is roughly
constant across the filter scale $R$.

The main advantage of the method proposed here is that the
signal-to-noise can be directly estimated from the data and thus
allows the determination of a significance level from the data
only. In addition, it can be easily calculated from a list of galaxy
images which contains positions and image ellipticities. For these
reasons, a signal-to-noise map can be routinely constructed for all
sufficiently deep wide-field images with good imaging quality. Since
wide-field cameras will become increasingly available, this method
will allow to perform a survey of mass concentrations selected by mass
only (in contrast to light)! A mass-selected sample of dark halos is
much more closely related to theories of structure formation in the
universe and can be directly compared with numerical simulations,
whereas optical or X-ray-selected samples of dark halos can be
compared with the results of numerical simulations of structure
formation only with additional assumptions concerning star formation
histories and efficiencies, feed-back processes (e.g., from supernova
explosions), or gas-dynamical processes. Results from lensing in
clusters like MS 1224 (Fahlman et al.\ts 1994) and Cl0939+4713 (Seitz et
al.\ts 1996) have yielded quite a broad range of
mass-to-light ratios, which implies that by `radiation-selected'
samples of dark halos those with low mass-to-light ratios will be
preferentially included. 

A first attempt to discover dark matter concentrations by weak lensing
has been successfully made by Fort et al.\ts (1996) by investigating
the shear field around high-redshift bright QSOs which are suspected
to be affected by the magnification bias. The method presented in this
paper will allow to extend the search for dark matter halos to a much
larger angular region on the sky.

Further improvements of the detection method are possible. For
example, in Sect.\ts 4.4 the effects of a PSF were not included in the
$S$-statistics explicitly, i.e., not attempt for correcting the image
ellipticities for a PSF was made. If the PSF is known and stable over
the filter size $R$, one can define in analogy to (50) an aperture
measure $m^\o$ in which the sum over the images is weighted according
to the size of the images; large images contribute a larger signal
than smaller images. The weights can again be optimized to yield the
largest signal-to-noise ratio. In addition, the detection of weak
shear from the surface brightness autocorrelation function (Van
Waerbeke et al.\ts 1996) provides an alternative and completely
independent means for measuring the shear. This will be the subject of
a future publication.

\SFB
\def\ref#1{\vskip1pt\noindent\hangindent=40pt\hangafter=1 {#1}\par}
\begpet
\sec{References}

\ref{Bartelmann, M.\ 1995, A\&A 303, 643.}
\ref{Bartelmann, M., Narayan, R., Seitz, S. \& Schneider, P.\ 1995, ApJ
(submitted). }
\ref{Bartelmann, M. \& Schneider, P.\ 1993, A\&A 271, 421.}
\ref{Bartelmann, M. \& Schneider, P.\ 1994, A\&A 284, 1.}
\ref{Benitez, N. \& Martinez-Gonzalez, E.\ 1995, ApJ 448, L89.}
\ref{Bonnet, H. \& Mellier, Y.\ 1995, A\&A 303, 331.}
\ref{Dalcanton, J.J., Canizares, C.R., Granados, A., Steidel, C.C. \&
Stocke, J.T.\ 1994, ApJ 424, 550.}
\ref{Fahlman, G., Kaiser, N., Squires, G. \& Woods, D.\ 1994, ApJ 437, 56.}
\ref{Fort, B. \& Mellier, Y.\ 1994, A\&AR 5, 239.}
\ref{Fort, B., Mellier, Y., Dantel-Fort, M., Bonnet, H. \& Kneib, J.-P.\
1996, A\&A, in press.}
\ref{Fugmann, W. 1990, A\&A 240, 11.}
\ref{Hutchings, J.B.\ 1995, AJ 109, 928.}
\ref{Kaiser, N.\ 1995, ApJ 439, L1.}
\ref{Kaiser, N. \& Squires, G.\ 1993, ApJ 404, 441.}
\ref{Kaiser, N., Squires, G. \& Broadhurst, T.\ 1995, ApJ 449, 460.}
\ref{Kaiser, N., Squires, G., Fahlman, G. \& Woods, D.\ 1994, in: {\it
Clusters of Galaxies}, eds. F.\ts Durret, A.\ts Mazure \& J.\ts Tran
Thanh Van, Editions Frontieres.}
\ref{Kochanek, C.S.\ 1993, ApJ 419, 12.}
\ref{Kochanek, C.S.\ 1995b, preprint (astro-ph/9510077).}
\ref{Maoz, D. \& Rix, H.-W.\ 1993, ApJ 416, 425.}
\ref{Miralda-Escud\'e, J. \& Babul, A.\ 1995, ApJ 449, 18.}
\ref{Rodrigues-Williams, L.L. \& Hogan, C.J.\ 1994, AJ 107, 451.}
\ref{Schneider, P.\ 1993, A\&A 279, 1.}
\ref{Schneider, P.\ 1995, A\&A 302, 639.}
\ref{Schneider, P., Ehlers, J. \& Falco, E.E.\ 1992, {\it Gravitational
lenses}, Springer: New York (SEF).}
\ref{Schneider, P. \& Seitz, C.\ 1995, A\&A 294, 411.}
\ref{Schramm, T. \& Kayser, R.\ 1995, A\&A 299, 1.}
\ref{Seitz, C. \& Schneider, P.\ 1995a, A\&A 297, 287.}
\ref{Seitz, C. \& Schneider, P.\ 1996a, A\&A (submitted).}
\ref{Seitz, C., Kneib, J.-P., Schneider, P. \& Seitz, S.\ 1996, A\&A
(submitted).} 
\ref{Seitz, S. \& Schneider, P.\ 1995b, A\&A 302, 9.}
\ref{Seitz, S. \& Schneider, P.\ 1996b, A\&A, in press.}
\ref{Squires, G. \& Kaiser, N.\ 1996, ApJ (submitted).}
\ref{Tyson, J.A., Valdes, F. \& Wenk, R.A. 1990, ApJ 349, L1.}
\ref{Van Waerbeke, L., Mellier, Y., Schneider, P., Fort, B. \& Mathez,
G.\ 1996, in preparation.}
\ref{Villumsen, J.V.\ 1995, MNRAS, submitted.}
\endpet

\vfill\eject\end